\documentclass[twocolumn,prl,superscriptaddress]{revtex4-1}
\usepackage{bm}
\usepackage[fleqn]{amsmath}
\usepackage{hyperref}
\usepackage{graphicx}
\usepackage{lipsum}
\usepackage{multirow}

\begin{document}

\title{Stacking domains and dislocation networks in marginally twisted bilayers of transition metal dichalcogenides}
\author{V.V. Enaldiev}
%\email{vova.enaldiev@gmail.com}
\address{National Graphene Institute, University of Manchester, Booth St. E. Manchester, M13 9PL, United Kingdom}
\address{School of Physics and Astronomy, University of Manchester, Oxford Road, Manchester, M13 9PL, United Kingdom}
\address{Kotel'nikov Institute of Radio-engineering and Electronics of the Russian Academy of Sciences, 11-7 Mokhovaya St, Moscow, 125009 Russia}
\author{V. Z\'olyomi}
\address{Hartree Centre, STFC Daresbury Laboratory, Daresbury, WA4 4AD, United Kingdom}
\address{National Graphene Institute, University of Manchester, Booth St. E. Manchester, M13 9PL, United Kingdom}
\address{School of Physics and Astronomy, University of Manchester, Oxford Road, Manchester, M13 9PL, United Kingdom}
\author{C. Yelgel}
\address{Recep Tayyip Erdogan University, Department of Electricity and Energy, Rize, 53100, Turkey}
\address{National Graphene Institute, University of Manchester, Booth St. E. Manchester, M13 9PL, United Kingdom}
\address{School of Physics and Astronomy, University of Manchester, Oxford Road, Manchester, M13 9PL, United Kingdom}
\author{S.J. Magorrian}
\address{National Graphene Institute, University of Manchester, Booth St. E. Manchester, M13 9PL, United Kingdom}
\address{School of Physics and Astronomy, University of Manchester, Oxford Road, Manchester, M13 9PL, United Kingdom}
\author{V.I. Fal'ko}
\address{National Graphene Institute, University of Manchester, Booth St. E. Manchester, M13 9PL, United Kingdom}
\address{School of Physics and Astronomy, University of Manchester, Oxford Road, Manchester, M13 9PL, United Kingdom}
\address{Henry Royce Institute for Advanced Materials, University of Manchester, Manchester, M13 9PL, United Kingdom}

\date{\today}

\begin{abstract}
We apply a multiscale modeling approach to study lattice reconstruction in marginally twisted bilayers of transition metal dichalcogenides (TMD). For this, we develop DFT-parametrized interpolation formulae for interlayer adhesion energies of MoSe$_2$, WSe$_2$, MoS$_2$, and WS$_2$, combine those with elasticity theory, and analyze the bilayer lattice relaxation into mesoscale domain structures. Paying particular attention to the inversion asymmetry of TMD monolayers, we show that 3R and 2H stacking domains, separated by a network of dislocations develop for twist angles $\theta^{\circ}<\theta^{\circ}_P\sim 2.5^{\circ}$ and $\theta^{\circ}<\theta^{\circ}_{AP}\sim 1^{\circ}$ for, respectively, bilayers with parallel (P) and antiparallel (AP) orientation of the monolayer unit cells and suggest how the domain structures would manifest itself in local probe scanning of marginally twisted P- and AP-bilayers.   
\end{abstract}

\maketitle

Layer-by-layer assembly of van der Waals (vdW) heterostructures of two-dimensional crystals  became a popular method of creating new hybrid materials \cite{novoselov20162d}. The underlying physics of optoelectronic properties of such systems includes the interlayer hybridization of electronic states of the layers and the superlattice effects, produced by the periodic moir\'e patterns characteristic to pairs of mutually twisted or slightly incommensurate lattices. Such effects have been extensively investigated in graphene-hBN heterostructures \cite{Ponomarenko2013,dean2013hofstadter,kim2017evidence,LeeScience2016,yankowitz2012emergence} and twisted graphene bilayers \cite{cao2018unconventional,cao2018correlated,Yankowitz2019}, and these studies have identified two distinct structural forms of bilayers. One corresponds to larger twist angles, $\theta$, and a stronger lattice mismatch, $\delta$, for which the periodic variation of local stacking of the atoms (moir\'e pattern with period $\ell=a/\sqrt{\delta^2+\theta^2}$) is due to the superposition of rigid crystalline lattices of the two layers \cite{Ponomarenko2013,dean2013hofstadter,kim2017evidence,LeeScience2016,yankowitz2012emergence,cao2018unconventional,cao2018correlated,Yankowitz2019}. The other, ''marginally twisted bilayers'' \cite{Berdyugin2019} regime is peculiar to a very small misalignment in graphene bilayers which reconstruct into large-area Bernal stacking domains, separated by networks of domain walls \cite{alden2013strain,yoo2019atomic,alden2013strain,yoo2019atomic,PhysRevB.92.155438,gargiulo2017structural,van2015relaxation,dai2016twisted,PhysRevB.96.075311,kerelskyArxiv2019}.

\begin{figure}[!t]
	\includegraphics[width=\columnwidth]{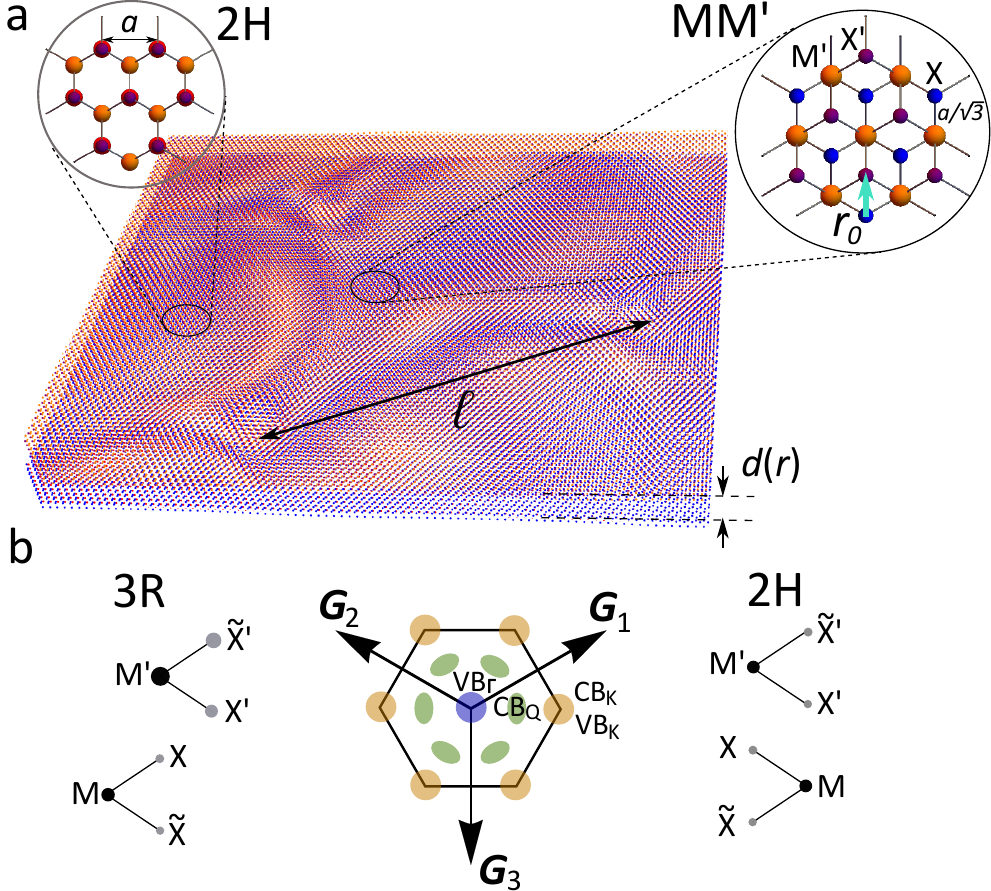}
	\caption{\label{Fig0} (a) Sketch of lattice relaxation across moir\'e supercell of  marginally twisted AP-bilayers ($\theta<\theta_{AP}$). Insets show local stacking configurations in 2H domain and MM$^\prime$ seed. X(X$^\prime$) and M(M$^\prime$) label chalcogen and metal atoms, respectively, in the bottom (top) layer; (b) the left/right panel reveals side-view of 3R/2H-stacked TMD bilayers with the size of dots reflecting the layer-asymmetry of the electronic states at the conduction band edge. Central panel: the first reciprocal lattice star and the first Brillouin zone of the TMD bilayers with marked conduction (CB$_Q$, CB$_K$) and valence band (VB$_\Gamma$, and VB$_K$) extrema. 
	}
\end{figure}

\begin{figure}[t]
	\includegraphics[width=1\columnwidth]{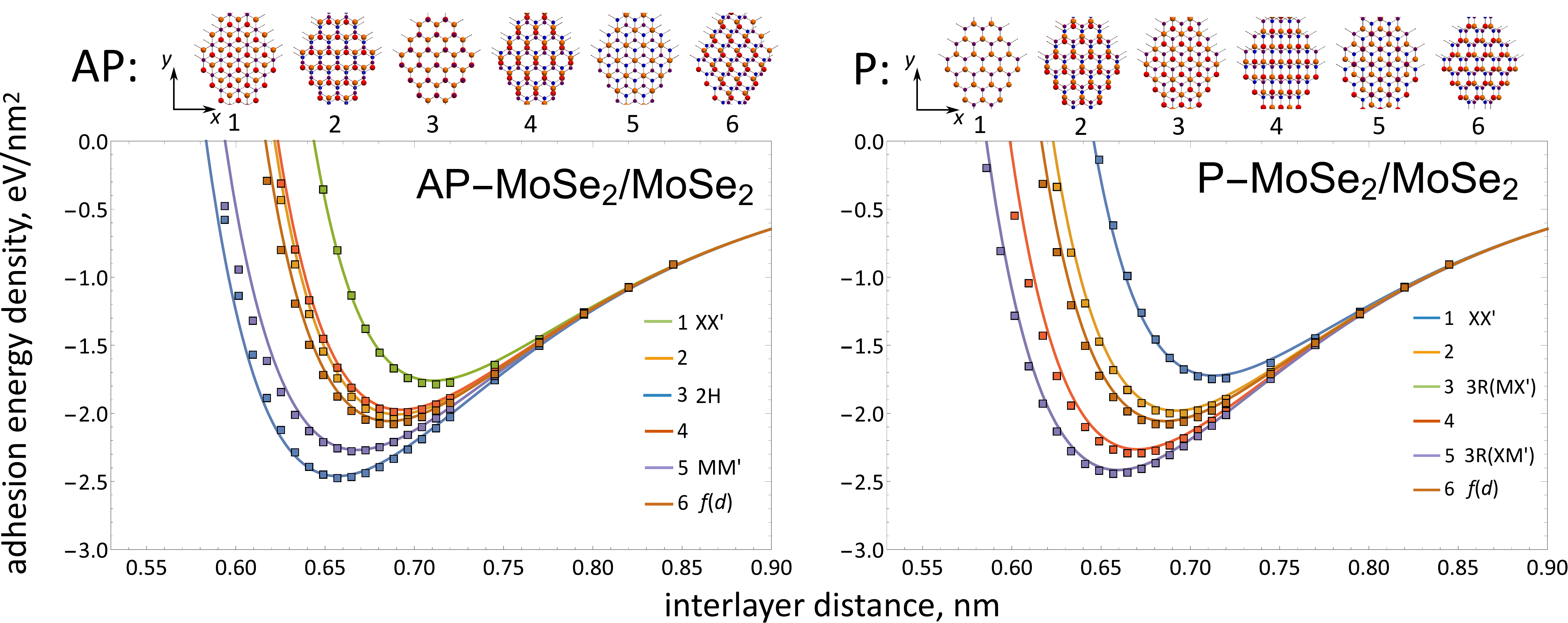}
	\caption{\label{FigSelenides} VdW-DFT data for MoSe$_2$/MoSe$_2$ bilayers and their fits using $W_{P/AP}$ in Eq.(\ref{adh_en}) (for WSe$_2$/WSe$_2$, MoSe$_2$/WSe$_2$, MoS$_2$/MoS$_2$, WS$_2$/WS$_2$, and MoS$_2$/WS$_2$ bilayers see SM \cite{SM}). Fitting parameters are listed in Table S1 in SM \cite{SM}.}
\end{figure}

Beyond graphene and hBN, moir\'e superlattice effects have been observed in MX$_2$/M$^\prime$X$^{\prime}_2$ heterostructures of transition metal dichalcogenides (TMD) \cite{kunstmann2018momentum,rivera2018interlayer,nayak2017probing,seyler2019signatures,tran2019evidence,jin2019observation,alexeev2019resonantly}, and it has been suggested theoretically \cite{PRLNaik,carr2018relaxation,naik2019kolmogorov} that twisted bilayers of TMD can undergo lattice reconstruction. Here, we determine parametric conditions for the formation of and the types of domain structures in twisted TMD homo- and heterobilayers, enriched by the lack of inversion symmetry of the individual 2D crystals. Different domain wall networks form for parallel- (P-bilayers) and antiparallel- (AP-bilayers) orientations of unit cells in the two layers (shown on top of Fig. \ref{FigSelenides}), and we find crossover angles,  $\theta^{\circ}_{AP}\sim 1.0^{\circ}$ and $\theta^{\circ}_{P}\sim 2.5^{\circ}$, for the marginality of the twist and, then, discuss how the resulting domain structures can be observed in scanning tunneling experiments.   

In this study we employ the multiscale modeling: a combination of density functional theory (DFT) leading to interpolation formulae for adhesion energy, $W_{P/AP}(\bm{r}_0,d)$, between the layers at a distance $d$ from each other and lateral offset $\bm{r}_0$, and elasticity theory for the lattice relaxation \cite{carr2018relaxation}. We perform this analysis for small misalignment angles, $\theta\ll 1$ (i.e $\theta^{\circ}<5^{\circ}$), and lattice mismatch $\delta\ll 1$. In this case energetics of local stacking can be described in terms of a lateral offset, \mbox{$\bm{r}_0\left(\bm{r}\right)=\theta \hat{z}\times\bm{r} + \delta\bm{r}+\bm{u}^{(t)} - \bm{u}^{(b)}$}, between two aligned commensurate lattices, which varies across the moir\'e supercell and includes the effect of local deformations, $\bm{u}^{(b/t)}(\bm{r})$, in the bottom/top layers. This multiscale approach enables us to overcome the system-size limitations of molecular dynamics 
simulations \cite{PRLNaik,naik2019kolmogorov}.
 
For adhesion energy, we use the form, $W_{P/AP}(\bm{r}_0,d)=\sum_n f^{(P/AP)}_n(d)e^{i\bm{G}_n\bm{r}_0}$, where $\bm{G}_n$ are the reciprocal lattice vectors of TMD. We truncate this sum at the first star of reciprocal lattice vectors, $\pm \bm{G}_{1,2,3}$, ($\left|\bm{G}_{1,2,3}\right|=G$, Fig. \ref{Fig0}) and set $\bm{r}_0=0$ at the XX$^\prime$ stacking for both P- and AP-bilayers. This choice -- together with the $D_{3h}$ lattice symmetry of TMD monolayers  -- suggests \footnote{This is because aligned P-oriented TMD bilayers with opposite in-plane offsets represent mirror copies of each other resulting in the same interaction value for $\pm\bm{r}_0$, whereas for AP orientation, the lack of inversion symmetry in each layer allows for both even and odd terms.} that $W_{P}=f(d) + f_1(d)\sum_{n=1}^{3}\cos\left(\bm{G}_n\bm{r}_0\right)$ and $W_{AP}=\widetilde{f}(d) + 
\sum_{n=1}^{3}\left[\widetilde{f}_1(d)\cos\left(\bm{G}_n\bm{r}_0\right) + \widetilde{g}_1(d)\sin\left(\bm{G}_n\bm{r}_0\right)\right] $. Then, we inspect the adhesion energies for various bilayers, computed using vdW-DFT with the optB88 functional \cite{SM} implemented in Quantum Espresso \cite{giannozzi2009quantum} for stacking configurations shown in Fig. \ref{FigSelenides}. For P-bilayers, the most energetically favorable are configurations MX$^\prime$ ($\bm{r}_0=(0,-a/\sqrt{3})$) and XM$^\prime$ ($\bm{r}_0=(0,a/\sqrt{3})$), which have equal energies \footnote{For MoSe$_2$/WSe$_2$ (Fig. \ref{FigSelenides}) and MoS$_2$/WS$_2$ (Fig. S1e,f in SM \cite{SM}) heterobilayers energies of twins (MX$^\prime$ and XM$^\prime$) of 3R-stacking configuration are almost the same.} and correspond to twins of a 3R bulk phase of a TMD. For AP-bilayers, 2H-stacking ($\bm{r}_0=(0,-a/\sqrt{3})$) has the lowest energy (in agreement with the 2H bulk phase of these TMDs), rather than  MM$^\prime$-stacking (configuration 5, $\bm{r}_0=(0,a/\sqrt{3})$) suggested in Ref. \cite{carr2018relaxation}. Configurations 6, \mbox{$\bm{r}_0=-a(1/3,1/\sqrt{3})$} and  \mbox{$\bm{r}_{0}=-(a/3,0)$}, are such that $W_{P}=f$ and $W_{AP}=\widetilde{f}$. The remaining two (2 and 4) have offsets $\bm{r}_0=(0, -a/2\sqrt{3})$ ($\bm{r}_0=(0, a/2\sqrt{3})$) and $\bm{r}_0=(0, a/2\sqrt{3})$ ($\bm{r}_0=(0,a\sqrt{3}/2)$) for AP(P)-bilayers. 

Using the DFT data for bilayers shown in Fig. \ref{FigSelenides} and S1 in SM \cite{SM}, we find the $d$-dependence of the factors $f,\widetilde{f}$, $f_1,\widetilde{f}_{1}$, and $\widetilde{g}_{1}$, and plot them in Fig. S2. By inspection, we find that $f\approx \widetilde{f}$ and $f_1\approx \widetilde{f}_1+\widetilde{g}_1$, over the broad interval of interlayer distances that covers the minima of $W_{P/AP}(\bm{r}_0,d)$, and, then, consider the following factors to make an informed choice of functions, $\widetilde{f}$, $\widetilde{f}_{1}$, and $\widetilde{g}_{1}$. (i) Coulomb potential of a lattice of $\pm q$ ions, whose potential decays exponentially, $\sum_{n}\alpha_n e^{-G_n d}e^{i \bm{G}_n\bm{r}_0}$, with the distance from their plane (\mbox{$\alpha_0=0$} is due to the electroneutrality of each layer). This suggests a choice of $\widetilde{g}_1=A_2e^{-Gd}$ for the first star of reciprocal lattice vectors $\pm \bm{G}_{1,2,3}$.  (ii)
The overlap of atomic orbitals, strongest for chalcogen atoms in the outer (top/bottom) sublattices in each layer is determined by the exponential decay of atomic wave functions away from the plane, \mbox{$|\psi(z)|^2\propto e^{-|z|/\rho}$}. Also, tunneling between the layers is suppressed for electrons with a larger in-plane wave number \cite{bistritzer2011}, \mbox{$\sum_{n}\beta_ne^{-d\sqrt{G_n^2+\rho^{-2}}}\cos\left(\bm{G}_n\bm{r}_0\right)$}, so that we choose $\widetilde{f}_1=A_1e^{-d\sqrt{G^2+\rho^{-2}}}$. (iii)
Finally, following the earlier studies \cite{PhysRevLett.96.073201} of vdW interaction of TMDs, which has a long-distance asymptotic, $\propto -C/d^4$,  we combine the short range repulsion and long-range vdW attraction into $\widetilde{f}=-\sum_{n=1}^{3}C_{n}/d^{4n}$. Then, we use   
{\setlength{\mathindent}{0cm}
\begin{multline}\label{adh_en}
W_{P/AP}(\bm{r}_0,d) = 
\sum_{n=1}^{3}\left[-\frac{C_{n}}{d^{4n}} + 
A_1 e^{-\sqrt{G^2+\rho^{-2}}d}\cos\left(\bm{G}_n\bm{r}_0\right)   
\right. \\ 
\left.
 + A_2 e^{-G d}\sin\left(\bm{G}_n\bm{r}_0+\varphi_{P/AP}\right)\right],  
\end{multline}}
with $\varphi_P=\pi/2$, $\varphi_{AP}=0$, and fit the values of parameters $C_{1,2,3}$, $A_{1,2}$, $\rho$ to the DFT data listed in Table S1 in SM \cite{SM}.  

Lattice reconstruction in bilayers happens when energy gain from the formation of favourable stacking overcomes elastic energy cost of strain produced by the local mutual adjustment of the two lattices, $U=\sum_{l={t,b}}\left[(\lambda_l/2)\left(u_{ii}^{(l)}\right)^{2} + \mu_l \left(u_{ij}^{(l)}\right)^{2}\right]$. Here, $\lambda_{t/b}$, $\mu_{t/b}$ and $u_{ij}^{(t/b)}=\frac{1}{2}(\partial_ju_{i}^{(t/b)}+\partial_iu_{j}^{(t/b)})$, are the first Lam\'e coefficient, shear modulus, and strain tensors related to the local in-plane displacements in top/bottom layer. Values of Young's moduli and Poisson ratios determining $\lambda$ and $\mu$ coefficients for TMD crystals under consideration are listed in Table S3 in SM\cite{SM}. We neglect the energy cost of flexural deformations \footnote{Up to room temperature the variation of interlayer distances, $\sqrt{\langle\delta d^2\rangle_T}$, due to thermal out-of-plane vibrations is by an order of magnitude less than the difference, $|d_{\rm 2H}-d_0|$, between the interlayer distance for 2H stacking and its value averaged over stacking configurations present in a moire supercell (see SM \cite{SM} section S4). In fact, we use the estimated value $\sqrt{\langle\delta d^2\rangle_{300K}}\approx 4\cdot10^{-3}\,$nm for the size of the symbols showing the DFT-computed adhesion energies in Fig. \ref{FigSelenides} to point out that the following analysis is applicable to all $T\leq 300$\,K}, see section S3 in SM \cite{SM} allowing for the out-of-plane bending of the layers towards the optimal interlayer distance, $d_{P/AP}(\bm{r}_0)$, for each offset $\bm{r}_0$. We describe the latter by expanding $W_{P/AP}$ into Taylor series about the minimum point, $d_0$, of the zeroth Fourier harmonic term, \mbox{$f(d)\approx f(d_0)+\varepsilon\left(d-d_0\right)^2$} (Table S1 in SM \cite{SM}), and, then, find $\bm{u}^{(t/b)}(\bm{r})$ that minimize energy,
\begin{eqnarray}\label{functional}
		\mathcal{E}=\int d^2\bm{r}\left\{U - \varepsilon Z_{P/AP}^2 + \right. \qquad\qquad\qquad\qquad\qquad\quad  \\ \sum_{n=1}^{3}\left[A_{1}e^{-\sqrt{G^2+\rho^{-2}}d_0}\cos\left(\bm{g}_n\bm{r} + \bm{G}_n[\bm{u}^{(t)}-\bm{u}^{(b)}]\right)+\right. \nonumber 
		 \\
		\left.\left.  A_2e^{-Gd_0}\sin\left(\bm{g}_n\bm{r} + \bm{G}_n[\bm{u}^{(t)}-\bm{u}^{(b)}]+\varphi_{P/AP}\right)\right]  \right\}; \nonumber\\
		Z_{P/AP}=\frac{1}{2\varepsilon}\left.\frac{\partial}{\partial d}\left[f(d)-W_{P/AP}(\bm{r},d)\right]\right|_{d=d_0}. \nonumber
\end{eqnarray}
Here, \mbox{$\bm{g}_{n}=\delta\bm{G}_{n}-\theta\hat{z}\times\bm{G}_{n}$} are reciprocal lattice vectors of moir\'e superlattice, which we also use to expand $\bm{u}^{(t/b)}(\bm{r})$ in Fourier harmonics up to the eightieth reciprocal space star. Then, we minimize $\cal{E}$ with respect to those Fourier amplitudes numerically and obtain the displacements in each layer of the reconstructed bilayer. Using this method, we study moir\'e structure with $\theta^{\circ}\geq 0.2^{\circ}$.

\begin{figure}[!t]
	\includegraphics[width=0.9\columnwidth]{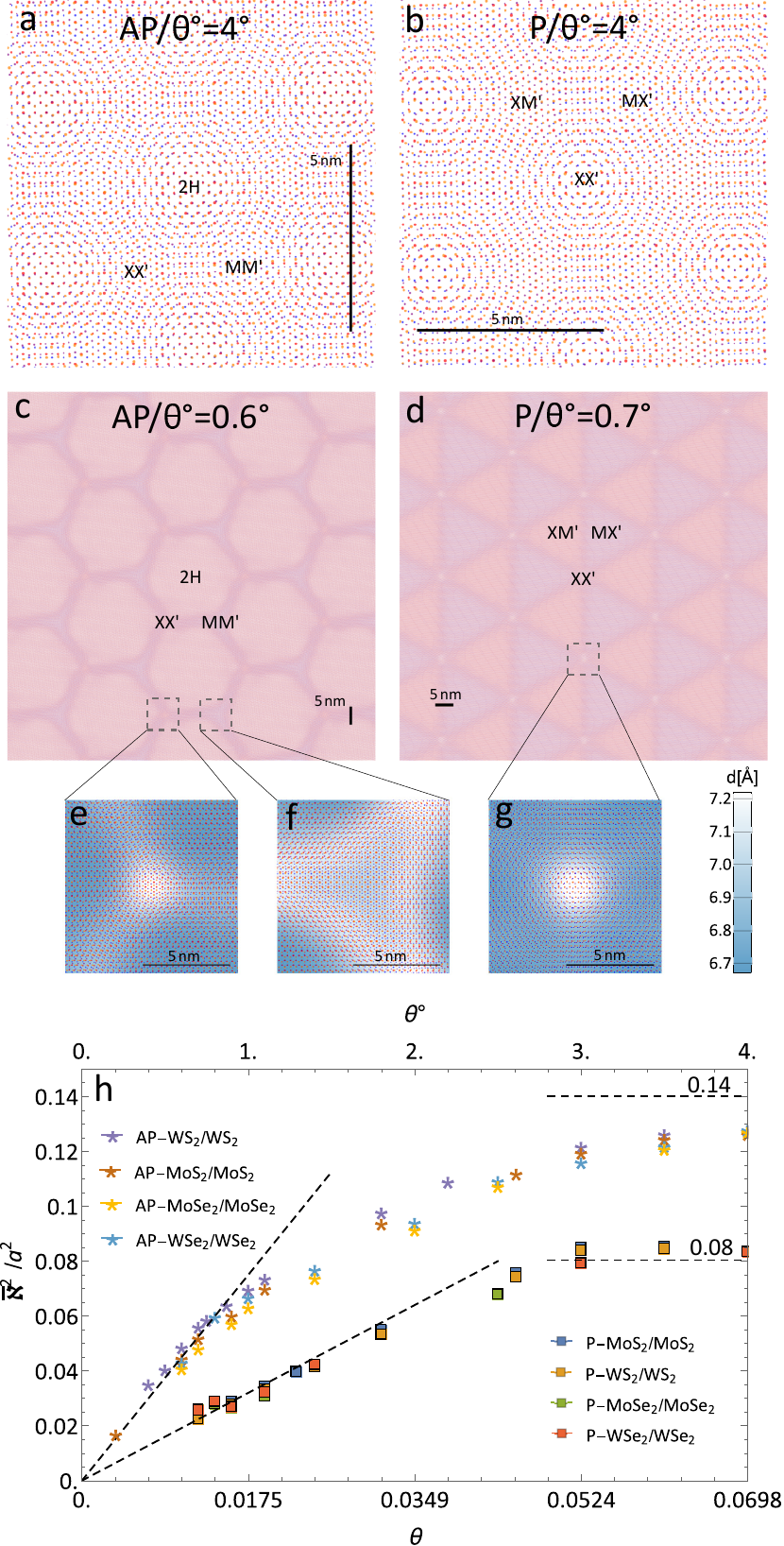}
	\caption{\label{FigReconstruction} (a-d) Reconstructed lattices of WSe$_2$/WSe$_2$ AP(P)-bilayer with $\theta^{\circ}=4^{\circ}>\theta_{AP/P}$ and $\,0.6^{\circ}$\,($0.7^{\circ}$)$\,<\theta_{AP/P}$ twist (for $\theta^{\circ}=0.2^{\circ}$ see SM \cite{SM}). For $\theta>\theta_{AP/P}$ the two layers behave as rigid, for marginally twisted bilayers (\mbox{$\theta\to 0$}) 2H for AP and 3R for P stacking domains emerge, separated by dislocations. (e-g) Intersections of dislocations, drawn over on the colour maps of the varied interlayer distance. (h) Crossover from rigid twist to the fully developed domain structure in P- and AP-bilayers, quantified using average value of parameter $\overline{\bm{\aleph}^2}$ described in the text.}
\end{figure}
	
Representative examples of lattice reconstructions in P- and AP-bilayers are shown in Fig. \ref{FigReconstruction}. For a larger angle, $\theta=4^{\circ}$ (Fig. \ref{FigReconstruction}a,b) adhesion-induced displacements are small and the two layers behave as rigid lattices. For $\theta^{\circ}<1^{\circ}$, twisted bilayers reconstruct into domain structures. For P-bilayers, Fig. \ref{FigReconstruction}d, each reconstructed moir\'e supercell comprises two equal area triangular domains of 3R(XM$^\prime$/MX$^\prime$) stacking, separated by partial dislocations, with XX$^\prime$ regions squeezed to the nodes on that partial dislocations network, Fig. \ref{FigReconstruction}g. For AP-bilayers, Fig. \ref{FigReconstruction}c, the reconstructed lattice features a honeycomb array of 2H domains separated by a dislocation network, where one half of the nodes are the seeds of the quasi-equilibrium MM$^\prime$ phase (stacking configuration AP-5). As a quantitative measure for domain formation, we use a lateral distance, $\bm{\aleph}$, between the closest metal and sulfur atoms in top and bottom layers (for ideal 2H and 3R domains, $\bm{\aleph}=0$) and compute its mean square, $\overline{\bm{\aleph}^2}/a^2$, over the supercell normalized by the TMD lattice constant, $a$. For rigidly rotated monolayers, $\overline{\bm{\aleph}_{AP}^2}\approx 0.14a^2$ and $\overline{\bm{\aleph}_P^2}\approx 0.08a^2$.
Upon the formation of 2H/3R domains, main contribution to $\overline{\bm{\aleph}^2}$ comes from domain boundaries so that the asymptotic behaviour, \mbox{$\overline{\bm{\aleph}}^2\propto 1/\ell\propto \theta$}, in Fig. \ref{FigReconstruction}h signals the formation of a domain structure at $\theta<\theta_{P}\approx 0.044\,(2.5^{\circ})$ and \mbox{$\theta<\theta_{AP}\approx 0.017\,(1^{\circ})$}. These quantitative estimates also explain why the molecular dynamics simulations performed in Refs. \cite{PRLNaik,naik2019kolmogorov} for $\theta^{\circ}>3^{\circ}$ failed to establish the full picture of the lattice reconstruction in H2DCs, having interpreted MM$^\prime$ areas in almost rigid AP-bilayers, Fig. \ref{FigReconstruction}a, as fully developed domains. Note that superlattice pattern -- perfect domains and the dislocation network -- also appear in TMD heterobilayers with the same chalcogens (WS$_2$/MoS$_2$ or WSe$_2$/MoSe$_2$), which have lattice mismatch \mbox{$\delta\lesssim 0.3\%$}, whereas bilayers with $\theta>\theta_{P/AP}$, or MS$_2$/M$^\prime$Se$_2$ heterostructures with $\delta\approx3.8\%$ behave as almost rigid crystals. 

\begin{figure}[t]
	\includegraphics[width=\columnwidth]{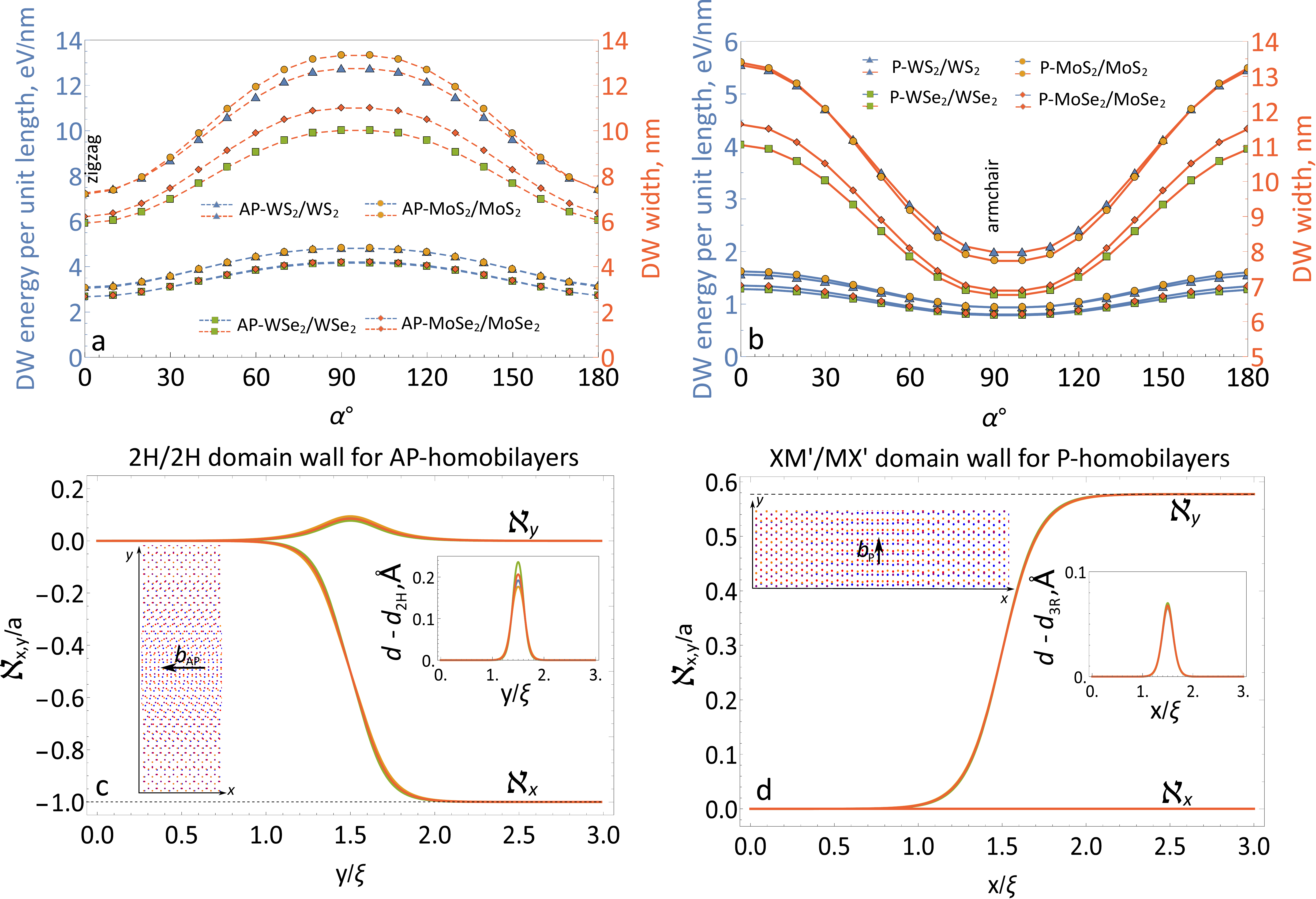}
	\caption{\label{FigDW} (a,b) Dependence of domain wall (DW) energy per unit length (blue) and width (red) on the angle between DW axis and zigzag direction ($x$-axis) in TMD layer, $\alpha$, for AP- and P-bilayers, respectively. (c) and (d) describe DW (dislocation) profile for AP- and P-bilayers, respectively. The calculated profiles for various TMDs (shown by different colours) almost coincide. Here, $\xi=a\sqrt{\mu/2A_1}\exp[d_0\sqrt{G^2+\rho^{-2}}/2]$. 
	}
\end{figure}

The formation of domain structures takes place when the energy gain from the expanded 2H (for AP) or 3R (for P) areas overcomes the energy cost of domain walls. The latter are nothing but dislocations: screw dislocations in 2H and partial screw dislocations in 3R bilayers. The properties of such linear defects, analysed using energy functional (\ref{functional}), are shown in Fig. \ref{FigDW}. Here, we set $\theta=0$ in Eq. (\ref{functional}) and replace $\bm{u}^{t}-\bm{u}^{b}$ by $\bm{\aleph}$, such that $\bm{\aleph}(-\infty)=0$ and $\bm{\aleph}(+\infty)=\bm{b}_{AP/P}$, where $\bm{b}_{AP}=a(-1,0)$ is a Burgers vector of a dislocation in 2H-TMD ($b_{AP}=a$), and $\bm{b}_{P}=a(0,1/\sqrt{3})$ is a partial dislocation Burgers vector in 3R bilayer ($b_{P}=a/\sqrt{3}$). A vector $\bm{n}=(\cos\alpha,\sin\alpha)$ determines the orientation of the dislocation line with respect to zigzag axis in the crystal. We find (see Fig. \ref{FigDW}) that the energy of a dislocation in 2H-homobilayers is the lowest and the cross-sectional width narrowest for a zigzag orientation of the defect line (screw dislocation); for 3R-homobilayers, the most favourable orientation of the partial dislocation axis is along the armchair direction (partial screw dislocation). These choices coincide with the orientations of domain boundaries shown in Fig. \ref{FigReconstruction}. In addition, in Fig. S8, we show the profile of edge dislocations that would form in perfectly aligned ($\theta<\delta$) heterobilayers MS$_2$/M$^\prime$S$_2$ and MSe$_2$/M$^\prime$Se$_2$ \footnote{The described domain structures and dislocation networks result from the fabrication-induced twist in a bilayer. As such, they are long-living quasi-equilibrium states, which may slowly relax upon annealing by dislocations escaping through the sample edges, resulting in larger-size domains of perfect 3R (for P) and 2H (for AP) stackings.}.

As a consequence of the domain formation, optoelectronic properties of marginally twisted bilayers would be dominated by the areas of 2H/3R stacking. While 2H bilayers have been deeply explored both theoretically \cite{PhysRevB.85.205302,PhysRevB.89.205311,PhysRevB.98.035411,PhysRevB.90.205420,chang2014,PhysRevB.92.205108,gong2013} and experimentally \cite{PhysRevLett.105.136805,suzuki2014valley,chernikov2015population,jones2014spin,wu2013electrical,jiang2014valley,lezama2015indirect,fallahazad2016shubnikov,PhysRevApplied.4.014002,PhysRevB.99.035443,PhysRevLett.123.117702,arora2018zeeman}, 3R bilayers were studied less \cite{PhysRevB.98.035408,zhang2018optical,yan2015stacking,suzuki2014valley,ubrig2017microscopic,PhysRevApplied.4.014002,wang2019unveiling,van2019stacking,yan2015stacking,shi20173r,mishina2015,zhao2016}, despite that the latter have interesting features resulting from their lack of inversion symmetry. Table \ref{tab_mos2} gives the atomic decompositions of the states at the edges of conduction and valence bands, computed using DFT for 3R bilayers (see SM for details \cite{SM}), highlighting the interlayer asymmetry of the states near band edges marked in Fig. \ref{Fig0}b. Such an asymmetry would make tunnelling characteristic of MX$^\prime$ and XM$^\prime$ domains different, leading to a manifestation of domain formation in conductive atomic force microscopy \cite{weston2019atomic} as well as to a linear Stark shift,  $(d_z({\rm CB_Q})-d_z({\rm VB_\Gamma}))E$, for the lowest-energy exciton transition in the bilayer, with the opposite sign of the shift in MX$'$ and M$'$X stacking domains \cite{Park_Falko}.
 
\begin{figure}[!t]
	\includegraphics[width=\columnwidth]{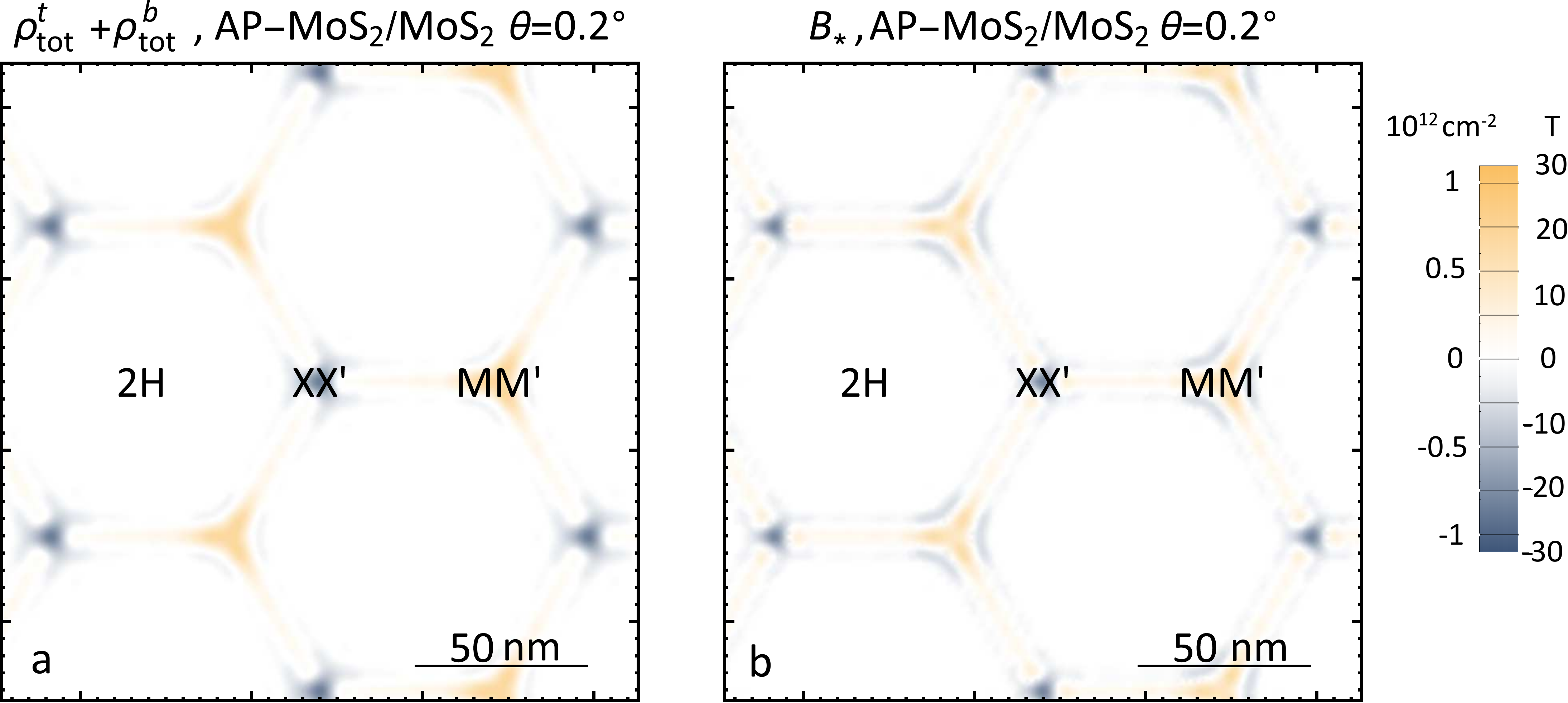}
	\caption{\label{Fig_B} (a,b) Distribution of piezoelectric charge density and pseudomagnetic field induced in AP bilayers MoS$_2$/MoS$_2$ ($\theta=0.2^{\circ}$). }
\end{figure}

\begin{table}
	\caption{
		{Orbital decomposition of DFT states at conduction (CB$_{\rm Q}$) and valence (VB$_{\Gamma}$) band edges of 3R homobilayers of MX$_2$ (as labelled in Fig. \ref{Fig0}b), the corresponding out-of-plane electric dipole moment, $d_z$, and ratio between tunneling currents into MX$^\prime$ and XM$^\prime$ stacking domains. \label{tab_mos2}
		}
	}
	\begin{tabular}{cc|cccccc|c|c}
		\hline\hline
		 & $\left|\psi\right|^2[\%]$	& $\widetilde{X}$ &M & X & X$^{\prime}$ & M$^\prime$ & $\widetilde{X}^\prime$ &$d_z$ (e\AA)  & I$_{MX^\prime}$\,$:$\,I$_{XM^\prime}$ \\
		\hline
		\multirow{2}{*}{\rotatebox[origin=c]{90}{MoS$_2$}} 	& CB$_{\mathrm{Q}}$& 4.2  & 30.6  & 3.3  & 5.5 & 49.6  &  6.8 & 0.73 & 1\,:\,1.6\\
		%&CB$_{\mathrm{K}}$& 0.0  & 0.0  & 0.0  & 2.4 &  95.3 & 2.4  & 3.07 & 0\,:\,1\\
		&VB$_{\Gamma}$& 10.6 & 41.8 & 0.9 & 0.7 & 36.0 & 10.0 & $-0.20$ & 1\,:\,1\\	
	%	&VB$_{\mathrm{K}}$&  4.7&  89.9&  4.7&  0.0&  0.5&  0.1& $-3.04$ &1\,:\,0\\
		\hline
		\multirow{2}{*}{\rotatebox[origin=c]{90}{MoSe$_2$}} &CB$_{\mathrm{Q}}$& 6.8  & 29.7 & 4.3 & 6.5 & 42.6 & 10.0 & 0.58 & 1\,:\,1.6\\
		%&CB$_{\mathrm{K}}$& 0.1 & 0.0  & 0.0 & 3.8 & 92.2  & 3.8 & 3.22 & 0\,:\,1\\
		&VB$_{\Gamma}$& 14.9 & 36.6  & 1.2 & 1.0  &31.9 & 14.3  & $-0.17$ & 1\,:\,1\\
		%&VB$_{\mathrm{K}}$& 7.5&  84.0& 7.5&  0.0&  0.7&  0.2& $-3.17$ & 1\,:\,0\\
		\hline
		\multirow{2}{*}{\rotatebox[origin=c]{90}{WS$_2$}} &CB$_{\mathrm{Q}}$& 4.3 & 28.6 & 4.0 & 6.8 & 48.9 & 7.4  & 0.81 & 1\,:\,1.7\\
		%&CB$_{\mathrm{K}}$& 0.0 & 0.0 & 0.0 & 1.8 & 96.3  & 1.8 & 3.09 & 0\,:\,1\\
		&VB$_{\Gamma}$& 10.3 & 41.5 & 1.6 & 1.3 & 35.5 & 9.9 & $-0.21$ & 1\,:\,1\\
		%&VB$_{\mathrm{K}}$& 6.5& 86.5 & 6.5& 0.0 & 0.5 & 0.2&  $-3.05$ & 1\,:\,0\\
		\hline
		\multirow{2}{*}{\rotatebox[origin=c]{90}{WSe$_2$}} &CB$_{\mathrm{Q}}$& 7.0 & 24.4 & 6.2 & 10.3 & 40.6 & 11.5  & 0.80 & 1\,:\,1.6\\
		%&CB$_{\mathrm{K}}$& 0.0 & 0.4 & 0.0 & 3.6 & 92.3 & 3.6 & 3.21 & 0\,:\,1\\
		&VB$_{\Gamma}$& 14.5 & 36.3 & 2.2 & 1.8 & 31.4 & 13.7 & $-0.19$ & 1\,:\,1\\
		%&VB$_{\mathrm{K}}$& 9.8& 79.0 & 9.9 & 0.7&  0.1&  0.4& $-3.16$ & 1\,:\,0\\
		\hline
		\hline
	\end{tabular}
\end{table}

Lacking inversion symmetry, TMD monolayers are piezoelectric crystals. The inhomogeneous strain concentrated in each layer at domain walls results in piezo-charges with density, $\rho^{(t/b)}_{piezo}=e_{11}\left[2\partial_xu_{xy}^{(t/b)}+\partial_y(u_{xx}^{(t/b)}-u_{yy}^{(t/b)})\right]$. In P-bilayers, the opposite signs of strain fields in the top and bottom layers leads to the mutual compensation of their piezoelectric charges. In AP-bilayers the inversion of the sign $e_{11}^{t}=-e_{11}^b$ of piezo-parameter in the two layers adds up charges, to the values shown in Fig. \ref{Fig_B}a, computed after taking into account partial screening of piezoelectric field by dielectric polarisability of TMD layers (see in SM \cite{SM} for details), with opposite signs in MM$'$ to XX$'$ corners of the domain wall network. Then, the MM$'$ regions and 2H/2H DWs will attract electrons in n-doped AP-homobilayers, giving rise to a larger tunnel current response than inside the 2H domains \cite{weston2019atomic}.  These spots of piezo-charges, with the values up to $\sim 10^{12}$\,cm$^{-2}$, can be viewed as ''quantum dots'' for electrons and holes in the vicinity of the band edges of the bilayers. The same areas are also the hot spots of pseudomagnetic fields ($B_*\propto \rho^{(t/b)}_{piezo}$, Fig. \ref{Fig_B}b), generated by inhomogeneous strain for electrons in multivalley semiconductors \cite{iordanskii1985}, which makes the dislocation network sites (MM$'$ and XX$'$ regions) interesting objects for optoelectronic studies. 

{\it Acknowledgements.} We thank R. Gorbachev, S. Haigh, and H. Park for fruitful discussions. This work has been supported by EPSRC grants EP/S019367/1, EP/S030719/1, EP/N010345/1; ERC Synergy Grant Hetero2D; Lloyd’s Register Foundation Nanotechnology grant; European Graphene Flagship Project, and EU Quantum Technology Flagship project 2D-SIPC. 

\bibliography{prlBibl}

%%%%%%%%%% Merge with supplemental materials %%%%%%%%%%
\pagebreak
%\widetext
\begin{center}
	\textbf{\large Supplemental Material for ''Stacking domains and dislocation networks in marginally twisted bilayers of transition metal dichalcogenides''}
\end{center}
%%%%%%%%%% Merge with supplemental materials %%%%%%%%%%
%%%%%%%%%% Prefix a "S" to all equations, figures, tables and reset the counter %%%%%%%%%%
\setcounter{equation}{0}
\setcounter{figure}{0}
\setcounter{table}{0}
\setcounter{page}{1}
\makeatletter
\renewcommand{\theequation}{S\arabic{equation}}
\renewcommand{\thefigure}{S\arabic{figure}}
\renewcommand{\thetable}{S\arabic{table}}
\renewcommand{\thesection}{S\arabic{section}}
%\renewcommand{\bibnumfmt}[1]{[S#1]}
%\renewcommand{\citenumfont}[1]{S#1}
%%%%%%%%%% Prefix a "S" to all equations, figures, tables and reset the counter %%%%%%%%%%

\section{Density functional theory (DFT) for adhesion energies in TMD bilayers and data analysis}

\begin{figure}[!htbp]
	\includegraphics[width=1\columnwidth]{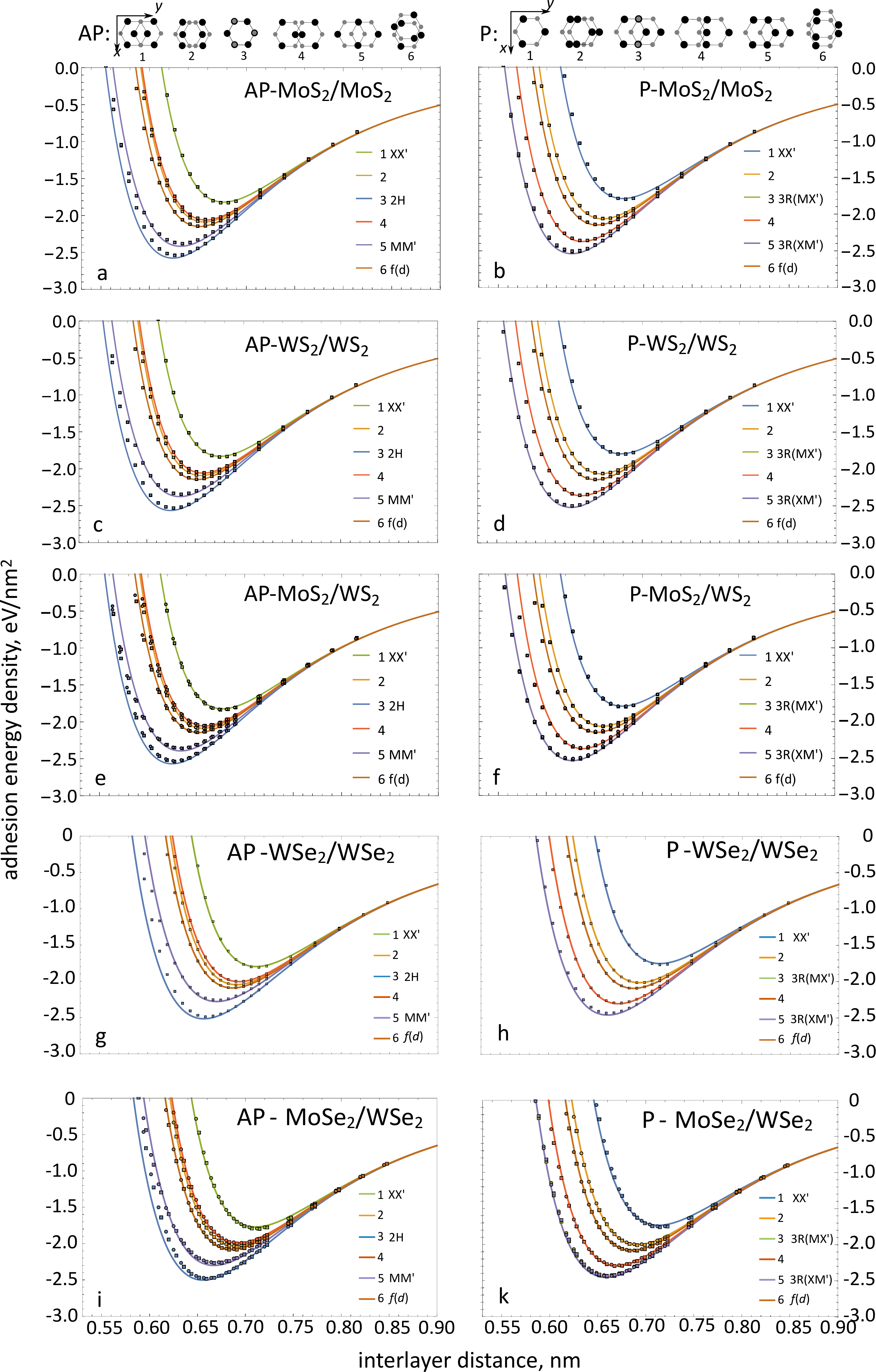}
	\caption{\label{FigMoS} Fitting of vdW-DFT data to $W_{P/AP}$ in Eq.\,(1) in the main text for AP/P-MoS$_2$/MoS$_2$, AP/P-WS$_2$/WS$_2$, AP/P-MoS$_2$/WS$_2$, AP/P-WSe$_2$/WSe$_2$, and AP/P-MoSe$_2$/WSe$_2$ bilayers.}
\end{figure}

In van der Waals-DFT (vdW-DFT) calculations of adhesion energies of TMD bilayers we neglected spin-orbit coupling, used a plane-wave cutoff of 816.34\,eV (60\,Ry), and kept the monolayer structure rigid, varying only interlayer distances and stacking ($\bm{r}_0$). In modelling of heterobilayers (MoSe$_2$/WSe$_2$ and MoS$_2$/WS$_2$) we fixed the lattice constants to be the same in both layers (to ensure local commensurability), while keeping  chalcogen-chalcogen distance in each monolayer equal to their individual monolayer values. For consistency, we performed these calculations using both (i) lattice constant of MoSe$_2$ (MoS$_2$) and (ii) lattice constant WSe$_2$ (WS$_2$) for both layers in MoSe$_2$/WSe$_2$ (MoS$_2$/WS$_2$) heterobilayer, comparing the differences. Such a comparison, Fig. \ref{FigMoS}e,f for sulphides (Fig. 2e,f of the main text for selenides), shows that these results are hardly distinguishable and their fitting using Eq.\,(1) produces almost identical parameters (the only noticeable difference is for the value of $A_2$, as shown for lateral lattice constant $a=0.329/0.328$\,nm ($a=0.316/0.3153$\,nm) in Table \ref{tab_Fit}).

\begin{figure}[!htbp]
	\includegraphics[width=1\columnwidth]{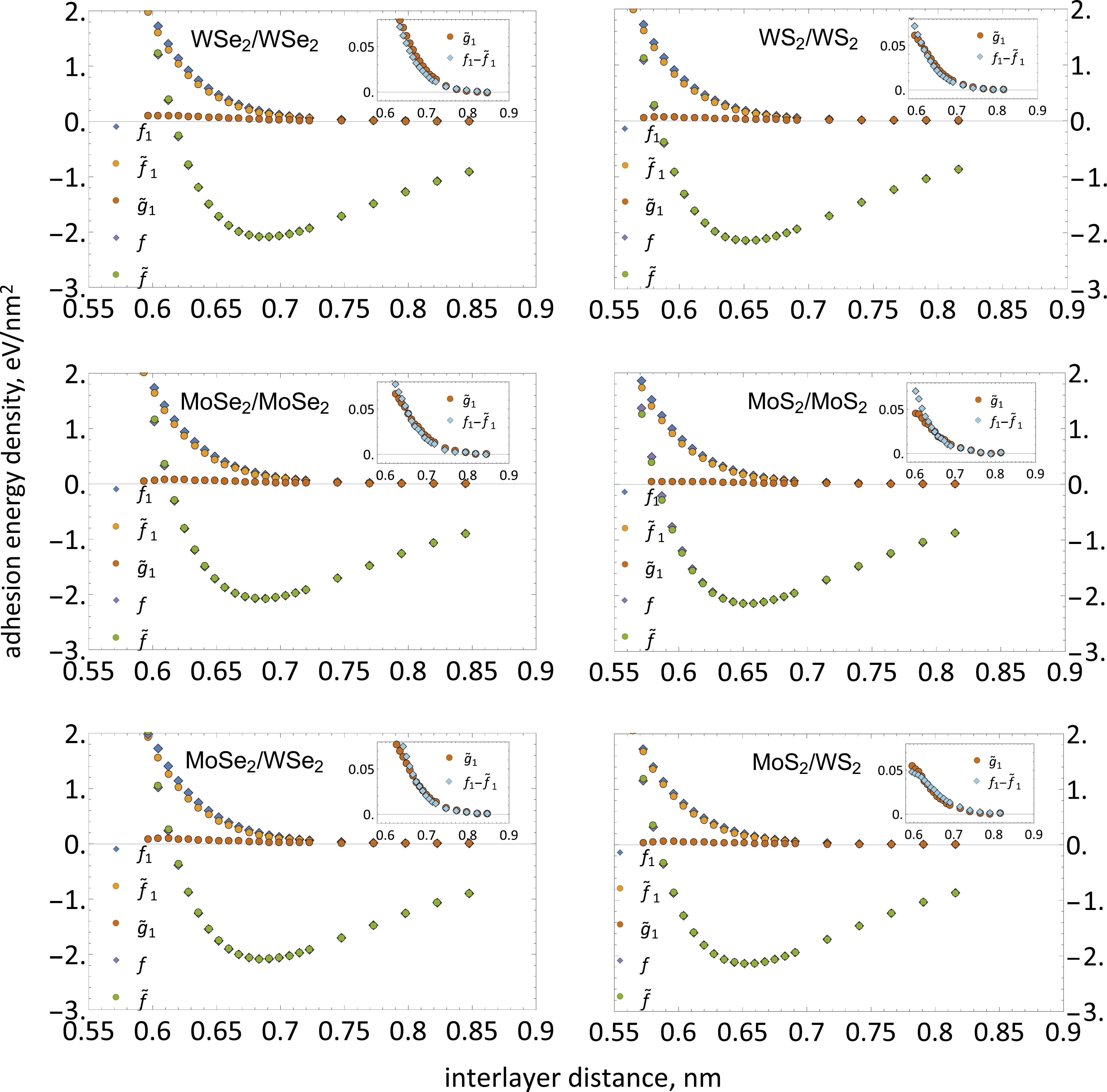}
	\caption{\label{Fig:enrg_func} Data-points for $f$, $\widetilde{f}$, $f_{1}$, $\widetilde{f}_{1}$ and $\widetilde{g}_1$ extracted from vdW-DFT data (Fig. \ref{FigMoS} and Fig. 2 in the main text) showing that $f=\widetilde{f}$ and $f_1=\widetilde{f}_1+\widetilde{g}_1$. }
\end{figure}

\begin{widetext}
	\begin{table}[!h]
		\caption{Fitting parameters for adhesion energy density for TMD bilayers.\label{tab_Fit}}
		%\centering
		%\footnotesize
		\begin{tabular}{|c|c|c|c||c|c|c||c|c|}
			%\hline
			\hline
			& $C_1$, & $C_2$, & $C_3$,  & \mbox{A$_{1}$}, & \mbox{A$_{2}$}, & $\rho$, & $d_0$& $\varepsilon$  \\ 
			& \mbox{eV$\cdot$nm$^2$} &\mbox{eV$\cdot$nm$^6$} &  \mbox{eV$\cdot$nm$^{10}$} &  eV/nm$^2$ &  eV/nm$^2$ & nm & nm & eV/nm$^4$\\
			\hline 
			WS$_2$/   &  \multirow{2}{*}{0.137976} & \multirow{2}{*}{0.159961} & \multirow{2}{*}{-0.020753} & \multirow{2}{*}{84571600} & \multirow{2}{*}{70214} &  \multirow{2}{*}{0.0495} & \multirow{2}{*}{0.65}  & \multirow{2}{*}{213}   \\
			WS$_2$&&&&&&&&\\
			\hline
			%& & & & & & \\
			MoS$_2$/  &  \multirow{2}{*}{0.134661} & \multirow{2}{*}{0.161589} & \multirow{2}{*}{-0.0209218} & \multirow{2}{*}{71928800} & \multirow{2}{*}{56411} &  \multirow{2}{*}{0.0496} & \multirow{2}{*}{0.65} & \multirow{2}{*}{214} \\
			MoS$_2$&&&&&&&&\\
			\hline
			WSe$_2$/ &  \multirow{2}{*}{0.148820} & \multirow{2}{*}{0.247806} & \multirow{2}{*}{-0.039458} & \multirow{2}{*}{121287200} & \multirow{2}{*}{110873} &  \multirow{2}{*}{0.0497} &\multirow{2}{*}{0.69} & \multirow{2}{*}{190} \\
			WSe$_2$ &&&&&&&&\\
			\hline
			MoSe$_2$/  &  \multirow{2}{*}{0.151965} & \multirow{2}{*}{0.233975} & \multirow{2}{*}{-0.0366911} & \multirow{2}{*}{96047400} & \multirow{2}{*}{81488} & \multirow{2}{*}{0.0506} & \multirow{2}{*}{0.68} & \multirow{2}{*}{189} \\
			MoSe$_2$ &&&&&&&&\\
			\hline
			MoS$_2$/  & \multirow{2}{*}{ 0.135693} & \multirow{2}{*}{0.162478} & \multirow{2}{*}{-0.0211387} & \multirow{2}{*}{79160000} & 63427/ &  \multirow{2}{*}{0.0492} & \multirow{2}{*}{0.65} & \multirow{2}{*}{214} \\
			WS$_2$ &&&&& 65461 &&&\\
			\hline
			MoSe$_2$/  &  \multirow{2}{*}{0.154394} & \multirow{2}{*}{0.236366} & \multirow{2}{*}{-0.0373244} & \multirow{2}{*}{77621500} & 84739/ &  \multirow{2}{*}{0.0520} & \multirow{2}{*}{0.69} & \multirow{2}{*}{189}\\
			WSe$_2$ &&&&& 88450 &&&\\
			\hline
			%\hline
		\end{tabular}
	\end{table}
	%\end{widetext}
	
	We use $W_{P/AP}$ (Eq. (1) in the main text) and DFT results for 2H-stacking to compute the frequencies of layer-breathing modes (LBM) for 2H and 3R TMD bilayers, and compare in Table \ref{tab_LBM} with those measured in Raman spectroscopy.
	
	%\begin{widetext}
	\begin{table}[!t]
		\caption{Phonon frequencies of layer-breathing modes (LBM), calculated using Eq.(1) of the main text, 2H vdW-DFT data, and compared to values measured using Raman spectroscopy\label{tab_LBM}. We used parameter $\varepsilon$  in configurations-averaged adhesion energy, $f(d)$, to estimate frequency of LBMs, $\omega_{\theta}^{LBM}$, in twisted bilayers.}
		%\centering
		\footnotesize
		\begin{tabular}{|c|c|c|c||c|c||c|}
			%\hline
			\hline
			\multirow{3}{*}{}  & $\omega^{LBM}_{2H}$, & $\omega^{LBM}_{2H}$, & $\omega^{LBM}_{2H}$,  &  $\omega^{LBM}_{3R}$, &  $\omega^{LBM}_{3R}$,  & $\omega^{LBM}_{\theta}$  \\
			& cm$^{-1}$  & cm$^{-1}$ & cm$^{-1}$ &  cm$^{-1}$ & cm$^{-1}$ &  cm$^{-1}$ \\
			&  exp. & 2H DFT & from Eq.(1) &  exp. & from Eq.(1) &  from Eq.(1) \\
			\hline
			\mbox{WS$_2$/WS$_2$}   & 33.8\cite{chen2015helicity} & 30.1 & 29.6 & & 29.4   & 28.3 \\
			\hline
			\multirow{3}{*}{\mbox{MoS$_2$/MoS$_2$}} & 41.6\cite{chen2015helicity,yan2015stacking}, & & & 41.6\cite{yan2015stacking} & &  \\
			& 41\cite{lin2018moire}, & 37.7 & 37.2 & 37\cite{van2019stacking} & 36.9 & 35.3  \\
			& 40\cite{zhang2013raman} & & & & & \\
			
			\hline
			MoSe$_2$/MoSe$_2$ & 34.3\cite{chen2015helicity} & 29.4 & 28.8 &  & 28.5 & 27.5 \\
			\hline
			WSe$_2$/WSe$_2$ & 29.1\cite{chen2015helicity}& 25.4 & 24.7 &  & 24.4 & 23.6  \\
			\hline
			MoS$_2$/WS$_2$& & 34.1 & 33.6 &  & 33.3 & 32  \\
			\hline
			MoSe$_2$/WSe$_2$ & & 27.6 & 25.9 & & 25.6  & 24.6   \\
			\hline
			%\hline 
		\end{tabular}
	\end{table}
\end{widetext}

\section{DFT band structure analysis for 3R and 2H bilayers}

The DFT calculations of 2H and 3R bilayer band structures were carried out in the local density approximation (LDA), as implemented in the VASP code \cite{PhysRevB.54.11169}, with spin-orbit coupling taken into account using projector augmented wave pseudopotentials. The cutoff energy for the plane-waves was set to 600eV, and the Brillouin zone sampled by a 12x12x1 grid. The bilayers are placed in a periodic three-dimensional box with a separation of 20\,\AA\, between repeated images to ensure no interaction between them. The structural parameters were taken from experimental measurements of bulk TMDC crystals \cite{schutte1987,bronsema1986}.

\section{Layer bending and thermal fluctuations due to layer-breathing modes in bilayers}

In this section, we study energy costs of layer bending and calculate magnitude of interlayer distance variation due to fluctuations caused by layer-breathing phonons at finite temperature. First, we recall that the interlayer distance variation caused by reconstruction,
\begin{eqnarray}\label{reconstruction_interlayer_distance}
Z_{AP/P}(\bm{r})=\frac{1}{2\varepsilon}\times\qquad\qquad\qquad\qquad\qquad\qquad\qquad\qquad\\
\sum_{n=1}^{3}\left[\sqrt{G^2+\rho^{-2}}A_{1}e^{-d_0\sqrt{G^2+\rho^{-2}}}\cos\left(\bm{g}_n\bm{r} + \bm{G}_n\bm{u}^{(tb)}\right)+ \nonumber \right. \\
\left. GA_{2}e^{-Gd_0}\sin\left(\bm{g}_n\bm{r} + \bm{G}_n\bm{u}^{(tb)}+\varphi_{AP/P}\right)\right],\nonumber
\end{eqnarray}
shown in Fig.3e-g of the main text (where $\bm{u}^{(tb)}=\bm{u}^{(t)}-\bm{u}^{(b)}$), does not exceed $\approx 0.6\,{\rm\AA}$, being much smaller than average interlayer distance $\approx 6.5\,{\rm\AA}$. Therefore, bending energy can be calculated as\cite{LL1986}:
\begin{eqnarray}\label{bending_functional}
E_{{\rm bend}}=\int d^2\bm{r}\left\{\sum_{l=t,b}\left[\frac{\kappa}{2}\left(\bm{\nabla_{\bm{r}}^2}Z_l\right)^2 + \frac{\rho \dot{Z}_l^2}{2}  \right.\right.\nonumber\qquad\qquad\\
\left.\left.+ 2\left(1-\sigma\right){\rm det}\left(\frac{\partial^2 Z_l}{\partial x_i\partial x_j}\right)\right]+ \varepsilon(\delta z_t - \delta z_b)^2\right\},\qquad\qquad 
\end{eqnarray} 
where $\kappa=E_{l}^{3D}d_0^3/12(1-\sigma^2)$, $\rho=2(m_M+2m_X)/a^2\sqrt{3}$ are flexural rigidity and surface density of each monolayer, respectively, $\bm{\nabla}_{\bm{r}}=(\partial_x,\partial_y)$ is 2D gradient operator, $Z_{t/b}=\pm Z_{P/AP}(\bm{r})/2 + \delta z_{t/b}(\bm{r},t)$ is local amendment to the average out-of-plane position of the top/bottom layer, $\pm d_0/2$, resulting from reconstruction and the layer-breathing phonons, respectively. Due to small magnitude of the interlayer distance variation caused by the layer-breathing phonons calculated below in this section, we estimate energy costs of monolayers' bending by taking zero temperature limit (i.e. $\delta z_{t/b}=0$). Then, we use ratio of the averaged over supercell bending and elastic ($U$) energies, 
\begin{equation}\label{ratioEU}
r=\frac{E_{{\rm bend}}}{U},
\end{equation}
to evaluate contribution of the former in energy cost of monolayers' deformations. For example, for marginally twisted AP- and P-MoS$_2$ bilayers at $\theta=0.6^{\circ}$ the ratios are
$r_{AP}=0.7\%$ and $r_{P}=0.5\%$, respectively, which validates our assumptions of negligible costs for the out-of-plane displacements made in the main text. Values of Young's moduli, $E^{3D}$, and Poisson ratios, $\sigma$, determining the bending and elastic energies for WS$_2$, MoS$_2$, WSe$_2$, and MoSe$_2$ crystals  are listed in Table \ref{tab_young_mod}.

\begin{figure}[!htbp]
	\includegraphics[width=1.0\columnwidth]{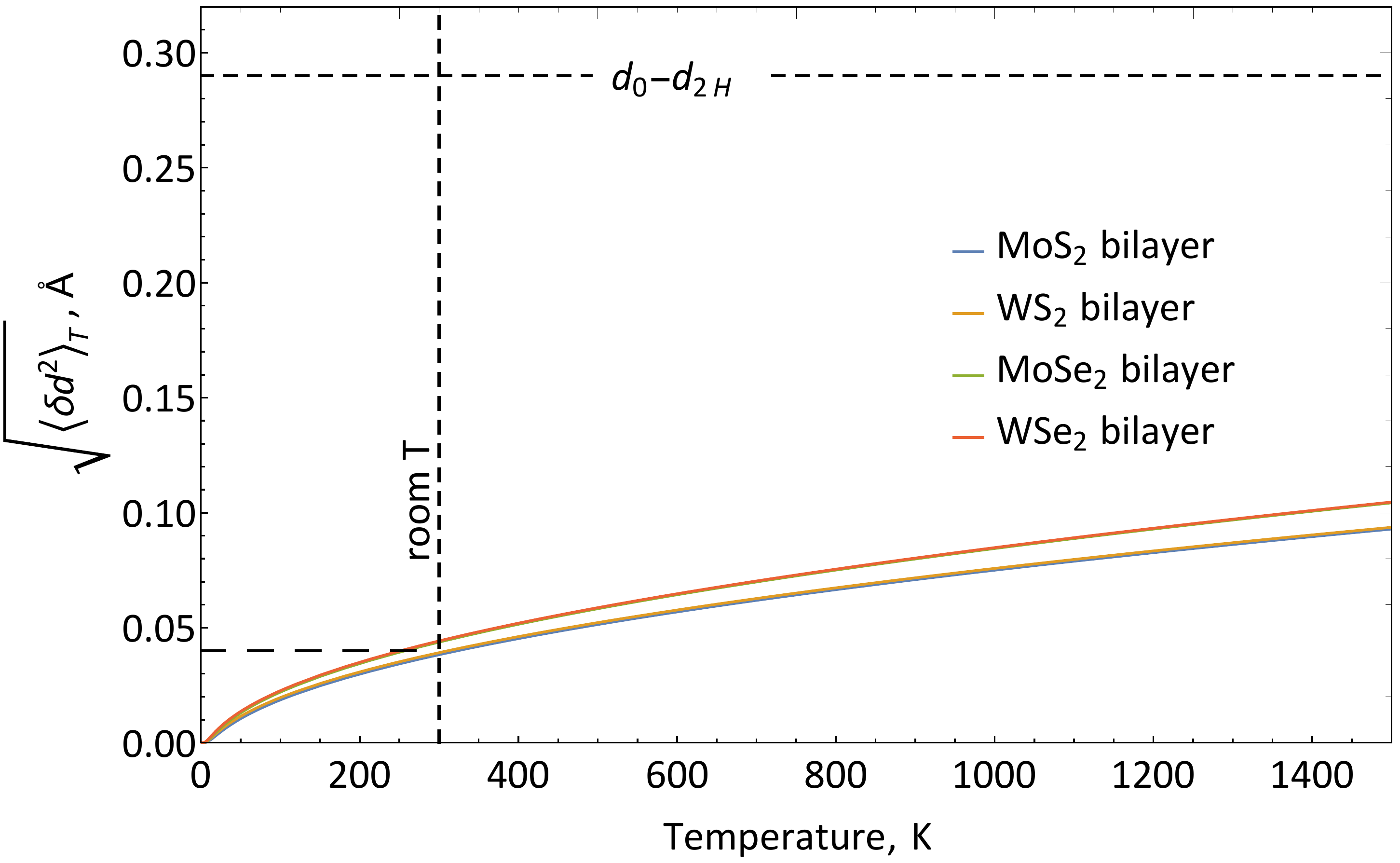}
	\caption{\label{Fig:ripple_phonon} Interlayer distance variations due to layer-breathing phonon modes in TMD bilayers, given by Eq. (\ref{avg_mean_root}). Dashed line shows difference in distances between optimal 2H-stacking and $d_0$, being extremum of the 6th configurations that corresponds to the adhesion energy averaged over moire supercell. }
\end{figure}

\begin{table}%[!t]
	\caption{Young's moduli and Poisson ratios used  for TMD crystals under consideration\cite{iguiniz2019,androulidakis2018tailoring}. Lam\'e coefficient and shear modulus incorporating in elastic energy, $U$, are expressed as follows: \mbox{$\lambda=E^{3D}d_0\sigma/(1+\sigma)(1-2\sigma)$}, $\mu=E^{3D}d_0/2(1+\sigma)$. \label{tab_young_mod} }
	%\centering
	%\footnotesize
	\begin{tabular}{|c|c|c|}
		%\hline
		\hline
		& $E^{3D}$, GPa & $\sigma$, Poisson ratio   \\
		\hline
		WS$_2$   & 270 & 0.21 \\
		\hline
		MoS$_2$  & 277 & 0.27 \\
		\hline
		MoSe$_2$ & 177 &  0.23\\
		\hline
		WSe$_2$ & 167 & 0.19  \\
		\hline
		%\hline 
	\end{tabular}
\end{table}

Further, we study thermal fluctuations of the interlayer distance, $\sqrt{\langle \delta d^2\rangle_T}$, produced by the layer-breathing phonons in the bilayers. We compare the variation with the difference between the optimal layer separation in 2H phase and $d_0$, which corresponds to the average interlayer separation over stacking configurations present in a moire supercell (corresponding to the 6-th configuration in Fig. \ref{FigMoS}). To describe these modes,  we separate relative $\delta d=\delta z_t-\delta z_b$ and centre mass $\delta z=\left(\delta z_t+\delta z_b\right)/2$ motions of the monolayers, and, neglecting the interlayer distance variation (\ref{reconstruction_interlayer_distance}) due to reconstruction, reduce Eq. (\ref{bending_functional}) to the following one
\begin{eqnarray}\label{bend_en2}
\widetilde{E}_{\rm bend}=\int d^{2}\bm{r}\left\{\frac{\rho \dot{\delta d}^2}{4} + \frac{\kappa}{4}\left(\bm{\nabla}^2_{\bm{r}} \delta d\right)^2 + \varepsilon (\delta d)^2\right\} \qquad\qquad \\
+\int d^{2}\bm{r}\left\{\rho \dot{\delta z}^2 + \kappa\left(\bm{\nabla}^2_{\bm{r}}\delta z\right)^2\right\}. \nonumber
\end{eqnarray}  
The first term in Eq. (\ref{bend_en2}) describes the layer-breathing mode, determining the phonon dispersion,
\begin{equation}\label{ripple_dispersion}
\omega_k=\sqrt{\frac{\kappa}{\rho}k^4 + \frac{4\varepsilon}{\rho}}\equiv\sqrt{\alpha k^4 + \omega^{LBM}}.
\end{equation} 
The variation of the interlayer distance, represented in terms of such phonons,
\begin{equation}
\delta d(\bm{r},t)=\sum_{\bm{k}}\sqrt{\frac{\hbar}{\omega_k\rho S}}\hat{z}\left[b_ke^{i\bm{kr}-i\omega_kt} + b_k^{\dagger}e^{-i\bm{kr}+i\omega_kt}\right], \nonumber
\end{equation}
enables us to estimate the effect of thermal vibrations, as
\begin{equation}\label{avg_mean_root}
\sqrt{\langle \delta d(\bm{r},t)^2\rangle_T}=\sqrt{\frac{\hbar}{4\pi \sqrt{\kappa \rho}}\int_{x_{min}}^{x_{max}}\frac{dx}{\sqrt{x^2-x^2_{min}}\left( e^{x} - 1 \right) }},
\end{equation}
where $x_{min}=\hbar\omega^{LBM}/T$, $x_{max}=\hbar\sqrt{\alpha G_1^4/16+\omega^{LBM}}/T$, $S$ is the bilayer area. In Fig. \ref{Fig:ripple_phonon} we present value of $\sqrt{\langle \delta d^2\rangle_T}$ computed using Eq. (\ref{avg_mean_root}) over a broad range of temperatures. It shows that variation of the interlayer distance due to thermally excited bending is much smaller than the difference, $d_0-d_{2H}$, so that relative energy difference between stacking configurations present across the moire supercell remain almost unchanged. In addition, we use the estimated value $\sqrt{\langle\delta d^2\rangle_{300K}}\approx0.04\,$\AA\, for the size of the symbols showing the DFT-computed adhesion energies in Fig. 2 of the main text to point out that the following analysis is applicable to all $T\leq 300$\,K.

Thus, our estimate shows that the domain networks in twisted TMD bilayers should be stable at room temperature. 

\section{Piezo-electric charges near domain boundaries}\label{sec3}

Piezoelectric tensor $e_{ijk}$ of TMD monolayers (which have $D_{3h}$ point symmetry group) has only one independent component, labeled by $e_{11}$ \cite{duerloo2012intrinsic,zhu2015observation}. For reference frame with the $y$-axis along one of the vertical mirror planes, the strain-induced charge polarization reads,
\begin{equation}
\bm{P}^{t/b}(\bm{r},z)=e^{t/b}_{11}\left(2u^{t/b}_{xy}, \left[u^{t/b}_{xx}-u^{t/b}_{yy}\right]\right)\delta\left(z-z_{t/b}\right).
\end{equation} 

Below, we use a reference frame with a positive direction of the $y$-axis coinciding with the in-plane projection of a vector from metal (M) to chalcogen atom (X) in the bottom layer. This leads to $e^{b}_{11}=-e^{t}_{11}>0$ for AP-bilayers, whereas for P-bilayers $e_{11}^b=e_{11}^t$. The polarization gives rise to piezo-charge densities in the two layers,
\begin{multline}\label{piezodensity}
\rho^{t/b}_{piezo}=-\bm{\nabla}_{\bm{r}}\cdot\bm{P}^{t/b}=\\
-e_{11}^{t/b}\left[2\partial_xu^{t/b}_{xy}+\partial_y\left(u^{t/b}_{xx}-u^{t/b}_{yy}\right)\right]\delta\left(z-z_{t/b}\right),
\end{multline}
that are of the same (opposite) sign in AP(P)-bilayers due to relation $u^{t}_{ij}(\bm{r})=-u^{b}_{ij}(\bm{r})$. 

\begin{figure}[!htbp]
	\includegraphics[width=1.0\linewidth]{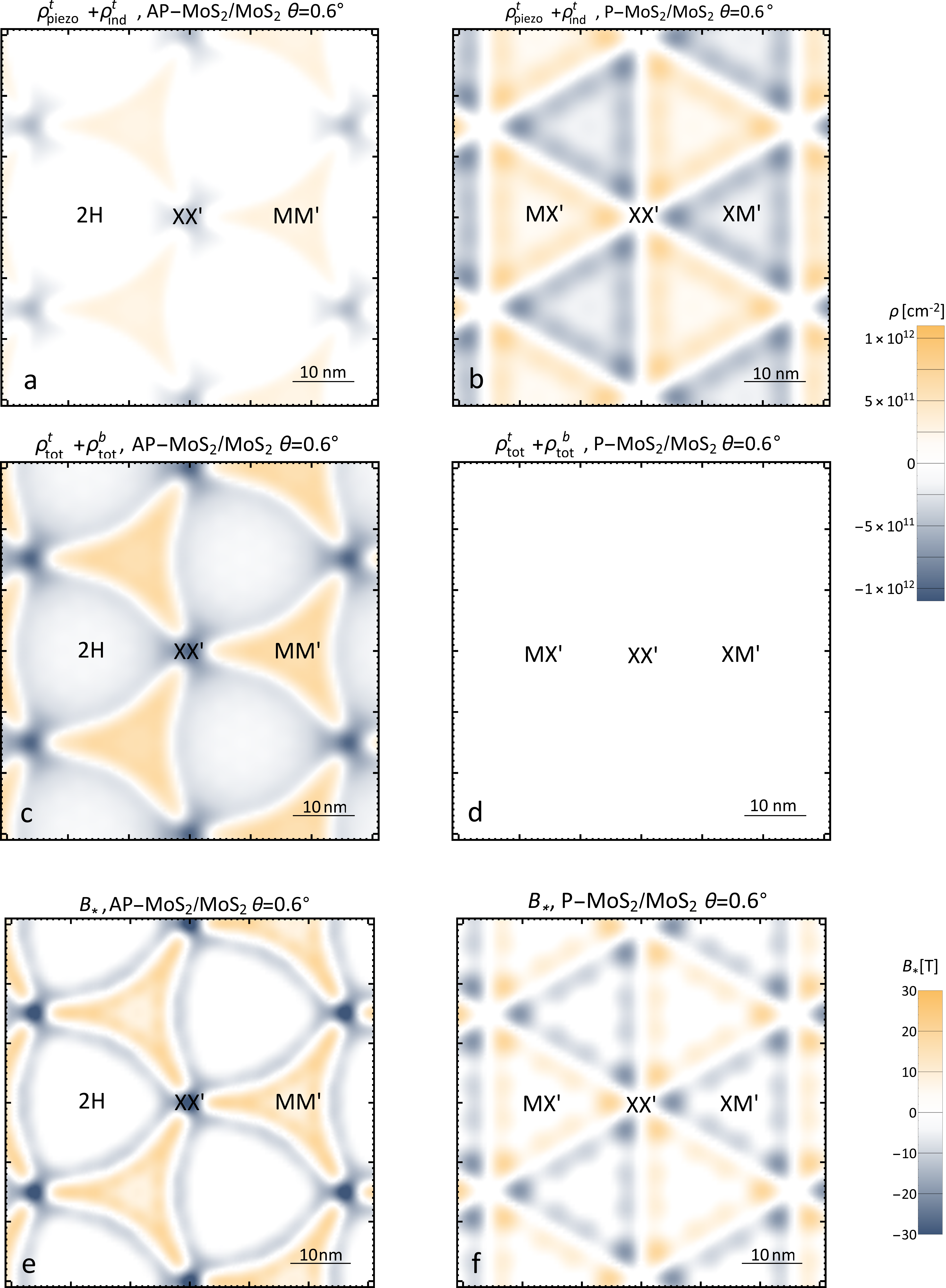}
	\caption{\label{Fig:2DchargeDensities} (a,b) distribution of total charge densities in the top layers of AP- and P-MoS$_2$ bilayers at $\theta=0.6^{\circ}$, respectively. (c,d) the same for sum of the densities in the two layers. Piezoelectric coefficient of MoS$_2$\cite{zhu2015observation}, $e_{11}=2.9\cdot 10^{-10} {\rm C\,m^{-1}}$ , in-plane dielectric permittivity of MoS$_2$\cite{laturia2018dielectric} , $\epsilon_{||}=16.3$. In P-bilayers the total charge densities are locally of opposite sign in top and bottom layers resulting in zero aggregated charge over whole system, shown in (d). (e,f) distribution of pseudomagnetic field in top layer of AP- and P-MoS$_2$ bilayers at $\theta=0.6^{\circ}$, respectively. An example of DW structure at even smaller misalignment angle is shown in Fig. \ref{Fig:APMoS2lattice}.}
\end{figure}

\begin{figure}[!htbp]
	\includegraphics[width=0.9\linewidth]{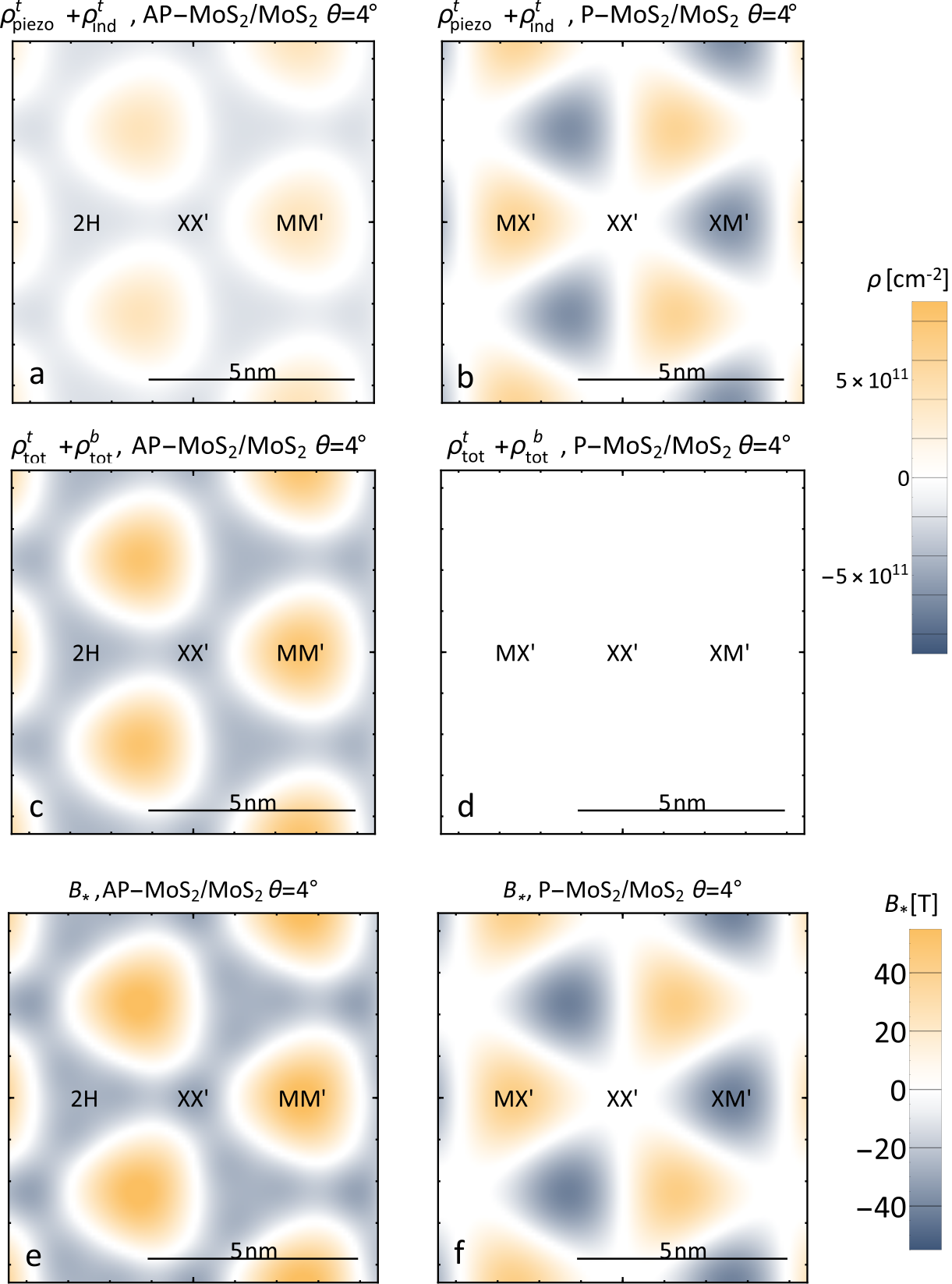}
	\caption{\label{Fig:2DchargeDensities4} same as in Fig. \ref{Fig:2DchargeDensities} but for $\theta=4^{\circ}$.}
\end{figure}

The total charge densities in each layer are sum of the piezo- and screening-induced charge densities. The latter emerges as a result of screening the former and is determined by the formula: $\rho^{t/b}_{ind}=\alpha^{t/b}_{2D}\delta\left(z-z_{t/b}\right)\bm{\nabla}^2_{\bm{r}}\varphi(\bm{r},z_{t/b})$, where $\varphi(\bm{r},z)$ is an electric potential produced by the piezo-charges, and $\alpha^{t/b}_{2D}$ is the in-plane 2D polarizability the top/bottom monolayer  related to the static in-plane dielectric permittivity of the TMD bulk crystal as: $\alpha^{t/b}_{2D}=d_0\left(\varepsilon_{||}-1\right)/4\pi$. 

To find the total charge densities, $\rho^{t/b}_{tot}=\rho^{t/b}_{piezo}+\rho^{t/b}_{ind}$, which include the above-mentioned screening, we solve the Poisson equation,  
\begin{equation}\label{Poisson_eq}
\left[\partial^2_{zz}+\bm{\nabla}^2_{\bm{r}}\right]\varphi=-4\pi\left(\rho^{t}_{tot}+\rho^{b}_{tot}\right), 
\end{equation}
by expanding the potential in Fourier series over the superlattice reciprocal vectors, $\varphi(\bm{r},z)=\sum_n\widetilde{\varphi}_n(z)e^{i\bm{g}_n\bm{r}}$, thus, using the already computed terms in Fourier series,  $\bm{u}^{t/b}(\bm{r})=\sum_n\bm{u}^{t/b}_ne^{i\bm{g}_n\bm{r}}$. Then, we find:
\begin{equation}\label{potential_harmonic}
\widetilde{\varphi}_n(z)=
\left\{
\begin{array}{l}
\varphi^t_{n} e^{-g_n(z-z_t)},\,z>z_t; \\
\varphi_{1n} e^{g_n(z-z_t)} + \varphi_{2n} e^{-g_n(z-z_b)}, \, z_b<z<z_t; \\
\varphi^b_n e^{g_n(z-z_b)}, \, z<z_b; 
\end{array}	
\right.
\end{equation}
where $g_n=|\bm{g}_n|$. From matching conditions for Eq. (\ref{Poisson_eq}), we obtain values of $\varphi^{t/b}_n$ on the top/bottom layer:
\begin{eqnarray}\label{potential_tb}
\varphi^{t/b}_n=\nonumber\qquad\qquad\qquad\qquad\qquad\qquad\qquad\qquad\qquad\qquad\qquad\quad \\
\frac{4\pi\sinh(g_n d)\left\{\rho_n^{t/b}+\rho_n^{b/t}\left[e^{g_nd} + 4\pi\alpha_{2D}^{t/b}g_n\sinh(g_nd)\right]\right\}}{g_n\left\{\prod_{l=t,b}\left[e^{g_nd}+4\pi\alpha_{2D}^{l}g_n\sinh(g_nd)\right]-1\right\}}, \qquad
\end{eqnarray}
where the Fourier amplitudes of piezo-charge density read as 
\begin{equation}\label{piezocharge_harmonics}
\rho_n^{t/b}=e_{11}^{t/b}\left[2g_{nx}g_{ny}u_{nx}^{t/b}+(g_{nx}^2-g_{ny}^2)u_{ny}^{t/b}\right].
\end{equation}

The total charge density in the top layer, $\rho^{t}_{tot}=\sum_n\left(\rho_n^{t}-4\pi g_n^2\alpha_{2D}^{t}\varphi_n^{t}\right)e^{i\bm{g}_n\bm{r}}$, and the total charge accumulated on two layers are shown on panels (a,b) and (c,d) in Fig. \ref{Fig:2DchargeDensities}\,(\ref{Fig:2DchargeDensities4}) for AP- and P-MoS$_2$ bilayers, respectively, with $\theta=0.6^{\circ}$ ($\theta=4^{\circ}$). In addition, panels e and f in Figs. \ref{Fig:2DchargeDensities} and \ref{Fig:2DchargeDensities4} demonstrate distributions of pseudomagnetic field in the top layer of twisted AP- and P-MoS$_2$, respectively (for details, see in the main text).  

\begin{figure}[!htbp]
	\includegraphics[width=0.8\linewidth]{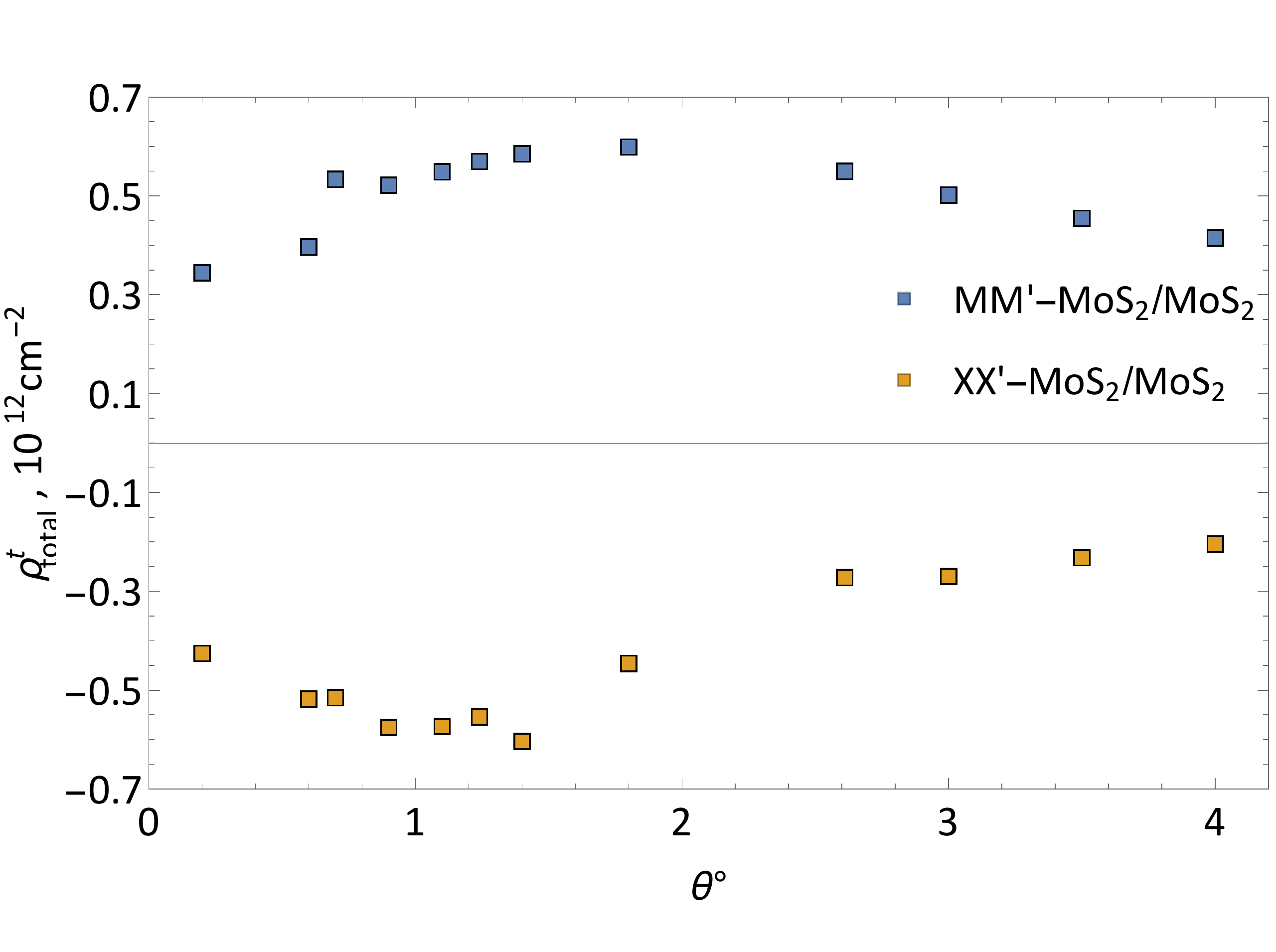}
	\caption{\label{Fig:angle_dependence_charge} Twist-angle-dependence of charge densities in the middle of XX$'$ and MM$'$ regions of top layer of AP-MoS$_2$ bilayer.}
\end{figure}

In Fig. \ref{Fig:angle_dependence_charge} we show twist-angle-dependence of the charge densities in the middle of XX$'$ and MM$'$ regions of the top layer of AP-MoS$_2$ bilayers. The absolute values of the charge densities in both regions are higher for smaller angles, saturating upon reaching the regime  of marginal twist corresponding to a clear 2H/2H domain wall formation. 

\section{Dislocations as Domain walls in TMD bilayers}

\begin{figure}[!tbp]
	\includegraphics[width=1.0\columnwidth]{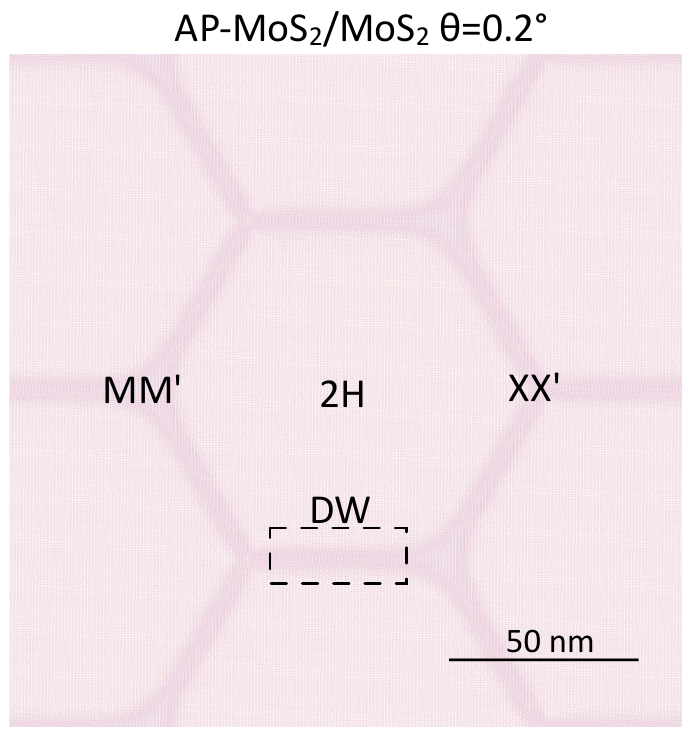}
	\caption{\label{Fig:APMoS2lattice} Reconstructed lattice of twisted AP-MoS$_2$ bilayers with twist angle $\theta=0.2^{\circ}$. Here, domain walls (DW) are screw dislocations, shown in Fig. 4c of the main text. Distribution of piezo-charge density and pseudomagnetic field, corresponding to twisted AP-MoS$_2$ bilayer with $\theta=0.2^{\circ}$, are shown in Fig. 5 of the main text. }
\end{figure}

\begin{figure}[!htbp]
	\includegraphics[width=1.0\columnwidth]{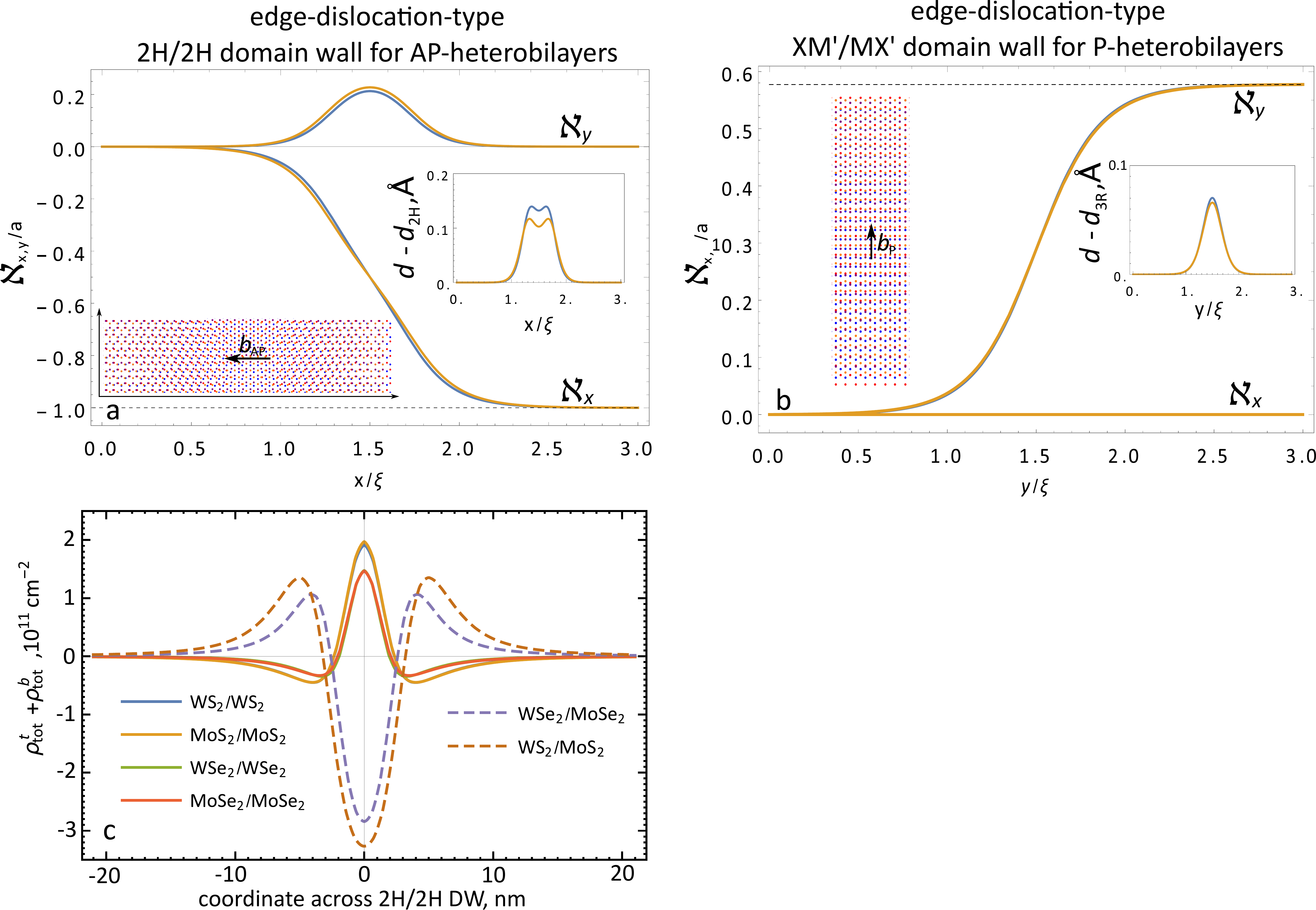}
	\caption{\label{Fig:DW_prop} (a,b) Profile of DW (edge dislocations) in AP- and P-heterobilayers (MoS$_2$/WS$_2$ and MoSe$_2$/WSe$_2$) with $\theta=0$,  $\xi=a\sqrt{\mu/2A_1}\exp[d_0\sqrt{G^2+\rho^{-2}}/2]$. (c) Piezo-charge density profiles across 2H/2H DW in homo- and heterobilayers of various TMDs.  }
\end{figure}

In Fig. \ref{Fig:APMoS2lattice} we show the domain structure in a marginally twisted bilayer with $\theta=0.2^{\circ}$. A large size of 2H stacking domains enables one to clearly indentify straight intervals of domain wall boundaries. In this section we discuss in detail domain walls (DW), emerging in marginally twisted bilayers. In Figs. \ref{Fig:DW_prop}(a,b) we present profiles of DW in AP- and P-heterobilayers (MoS$_2$/WS$_2$ and MoSe$_2$/WSe$_2$), respectively, with zero twist angle, which were calculated similar to those discussed in the main text. Note that, for heterobilayers with $\theta=0^{\circ}$, the DW are analogous to edge dislocations in layered crystal, in contrast to screw dislocations for twisted homobilayers.   

Using the calculated displacements for the 2H/2H DWs in hetero- and homobilayers, we computed corresponding profiles of piezo-charge densities across the DW in total for two layers, shown in Figs. \ref{Fig:DW_prop}c, taking into account dielectric screening considered in section \ref{sec3}. As the piezo-charge distributions are only determined by the second derivative of $\aleph_y$, they have opposite signs for screw dislocations ($\rho^{screw}_{piezo}=-e_{11}\partial^2_y\aleph_y(y)$) in twisted bilayers and edge dislocations ($\rho^{edge}_{piezo}=e_{11}\partial^2_x\aleph_y(x)$) in aligned heterobilayers. 

\end{document}